\pdfoutput=1

\documentclass[%
    aps                ,   
    pra                ,   
    reprint            ,   
    showpacs           ,   
    superscriptaddress ,   
    nofootinbib            
]{revtex4-2}
\usepackage[T1]{fontenc}
\usepackage[utf8]{inputenc}
\usepackage[english]{babel}

\usepackage{amsmath}
\usepackage{amssymb}
\usepackage{braket}
\usepackage{accents}
\DeclareMathOperator{\Tr}{Tr}
\DeclareMathOperator{\Imag}{Im}
\DeclareMathOperator{\Real}{Re}
\newlength{\dhatheight}
\newcommand{\doublehat}[1]{%
    \settoheight{\dhatheight}{\ensuremath{\hat{#1}}}%
    \addtolength{\dhatheight}{-0.35ex}%
    \hat{\vphantom{\rule{1pt}{\dhatheight}}%
    \smash{\hat{#1}}}}

\usepackage[hyperref]{xcolor}
\usepackage{graphicx}
\usepackage[export]{adjustbox}
\graphicspath{{figures/PNG/}{figures/PDF/}{figures/}}

\usepackage[
    unicode           = true  ,
    plainpages        = false , 
    pdfpagelabels     = true  , 
    bookmarks         = true  ,
    bookmarksnumbered = true  ,
    bookmarksopen     = true  ,
    breaklinks        = true  ,
    backref           = false ,
    colorlinks        = true  ,
    linkcolor         = blue  ,
    urlcolor          = blue  ,
    citecolor         = red   ,
    anchorcolor       = green ,
    hyperindex        = true  ,
    linktocpage       = true  ,
    hyperfigures      = true
]{hyperref}
\hypersetup{
    linkcolor         = [RGB]{000 114 189} ,  
    urlcolor          = [RGB]{000 114 189} ,  
    citecolor         = [RGB]{217 083 025} ,  
    anchorcolor       = [RGB]{119 171 048} ,  
    pdftitle={Signatures of Self-Trapping in the Driven-Dissipative Bose-Hubbard Dimer},
    pdfauthor={Matteo Seclì <matteo.secli@sissa.it>}
}

\begin{document}

\title{Signatures of Self-Trapping in the Driven-Dissipative Bose-Hubbard Dimer}

\author{Matteo Seclì}
\email[]{matteo.secli@sissa.it}
\affiliation{International School for Advanced Studies (SISSA), Via Bonomea 265, I-34136 Trieste, Italy}

\author{Massimo Capone}
\affiliation{International School for Advanced Studies (SISSA), Via Bonomea 265, I-34136 Trieste, Italy}
\affiliation{CNR-IOM Democritos, Via Bonomea 265, I-34136 Trieste, Italy}

\author{Marco Schirò}
\thanks{On Leave from: Institut de Physique Théorique, Université Paris Saclay, CNRS, CEA, F-91191 Gif-sur-Yvette, France}
\affiliation{JEIP, USR 3573 CNRS, Collége de France, PSL Research University, 11 Place Marcelin Berthelot, 75321 Paris Cedex 05, France}

\date{May 13, 2021}

\begin{abstract}
We investigate signatures of a self-trapping transition in the driven-dissipative Bose Hubbard dimer, in presence of incoherent pump and single-particle losses. For fully symmetric couplings the stationary state density matrix is independent of any Hamiltonian parameter, and cannot therefore capture the competition between hopping-induced delocalization and the interaction-dominated self-trapping regime. We focus instead on the exact quantum dynamics of the particle imbalance after the system is prepared in a variety of initial states, and on the frequency-resolved spectral properties of the steady state, as encoded in the single-particle Green's functions. We find clear signatures of a localization-delocalization crossover as a function of hopping to interaction ratio. We further show that a finite a pump-loss asymmetry restores a delocalization crossover in the steady-state imbalance and leads to a finite intra-dimer dissipation.
\end{abstract}

\maketitle

\section{Introduction}

Recent years have seen an increase of interest in open Markovian quantum systems, which describe a number of experimental platforms for quantum information processing and quantum simulation, both in the realm of atomic physics and quantum optics as well as in the solid state framework. Among these we can mention for example cavity QED experiments~\cite{Raimond2001} and their analogue with superconducting circuits~\cite{Blais2021}. Here the basic degrees of freedom, photons and qubits, are inevitably exposed to dissipative processes such as losses and decoherence induced by the environment. 
The quantum dynamics of Markovian systems is described theoretically within the framework of a Lindblad master equation which encodes the competition between coherent (Hamiltonian) evolution and dissipative processes described by a set of jump operators~\cite{Breuer2007}. Out of this competition one can expect non-trivial stationary states and dynamical behavior to emerge, leading to novel dissipative phase transitions~\cite{Kessler2012,Minganti2018}, both in small systems made by few quantum non-linear oscillators~\cite{Carmichael2015,Casteels2016} as well as in larger arrays~\cite{LeBoite2013,Schiro2016,Vicentini2018,Biella2017,Scarlatella2019,Landa2020,Landa2020a}. 

An intriguing question which has recently attracted large interest is to understand what kind of dynamical phenomena can arise in these Markovian quantum systems and their relationship with analogous phenomena in the field of classical non-linear dynamical systems in presence of non-linearities, noise and dissipation~\cite{Cross1993}.

A prototype example in this context is provided by the driven-dissipative Bose-Hubbard dimer (BHD), which can be seen as a toy model of strongly correlated open Markovian quantum systems since it encodes the basic competition between local dissipative processes, interactions and non-local coherent hopping processes.

Besides its paradigmatic relevance, the driven-dissipative BHD has also been realized experimentally in a variety of quantum light-matter platforms, including superconducting circuits~\cite{Raftery2014,Eichler2014} and semiconductor microcavities~\cite{Lagoudakis2010,Galbiati2012,Abbarchi2013} and photonic crystals~\cite{Hamel2015,Marconi2020}.

In the closed isolated case, corresponding to a purely conservative Hamiltonian evolution, the BHD has been extensively studied, in particular its self-trapping, or localization-delocalization~\cite{Smerzi1997,Pitaevskii2001,Polkovnikov2002,Albiez2005,Trujillo-Martinez2009,Venumadhav2010,Pudlik2013}. Here, an initial imbalance of particles between the two sites of the dimer is either rapidly redistributed by hopping processes leading to an homogeneous configuration or conserved indefinitely, leading to a self-trapped state below a critical ratio between hopping and interaction. This transition corresponds to a spontaneous breaking of the reflection symmetry between the two sites of the dimer. Open-Markovian extensions of the BHD have been mostly focused on the coherently driven case~\cite{Liew2010,Bamba2011,Eichler2014,Casteels2017,Seibold2020} or, in the case of the related Jaynes-Cummings Dimer model~\cite{Schmidt2010}, the purely dissipative case in absence of any external pumping.

In this work we theoretically study the driven-dissipative BHD in presence of single-particle losses and incoherent single-particle drive. This case is somewhat peculiar, since it is known that for a perfectly symmetric model the stationary state of the problem is completely independent of Hamiltonian parameters and only set by the ratio between pump and losses \cite{Lebreuilly2016}, so it cannot contain any signature of a putative delocalization transition. In order to explore the competition between hopping and interactions in a dissipative setting one has therefore to go beyond the analysis of steady-state observables and focus instead on response functions, or to consider an asymmetry between the two sites of the dimer. 

In particular we prepare the system in different initial states and follow the exact quantum dynamics of the model, characterizing also the properties of the stationary state reached at long times. Furthermore we focus on the spectral properties of the BHD as encoded in the Green's functions which for open-Markovian quantum system, much like their closed system counterpart, contain rich insights on the structure of the single-particle excitations around the stationary state.

The paper is organized as follows. In Sec.~\ref{sec:model} we introduce the BHD model and briefly review some of its properties, while in Sec.~\ref{sec:methods} we present details on its numerical solution. In Sec.~\ref{sec:semiclassical_dynamics_recap} we review the known results about the semiclassical limit and the self-trapping transition in the isolated and dissipative cases. Our results for the quantum dynamics in the symmetric pumping regime are discussed in Sec.~\ref{sec:quantum_time_dynamics}, while those for finite pump/loss asymmetry in Sec.~\ref{sec:res_quantum_ss}. In Sec.~\ref{sec:res_gf} we present results for the Green's functions of the BHD, while Sec.~\ref{sec:conclusions} is devoted to conclusions.

\section{The model}
\label{sec:model}

We start by considering the Hamiltonian of a Bose-Hubbard dimer (BHD). The model is a paradigmatic interacting lattice model which can be realized in a number of platforms. Our implementation including pumping and losses is naturally realized using optical cavities (see also Sec.~\ref{sec:discussion}). For this reason in the following we will refer to the two lattice sites as cavities and to the bosonic degrees of freedom involved in the physics as photons. The Hamiltonian reads
\begin{align}
    \hat{H} = 
    \omega_0 \left(\hat{n}_L  + \hat{n}_R\right)
    &+ U \Big( \hat{n}_L\hat{n}_L + \hat{n}_R\hat{n}_R \Big) \nonumber \\
    &+ J \Big( \hat{a}_L^{\dagger}\hat{a}_R + \hat{a}_R^{\dagger}\hat{a}_L \Big),
    \label{eq:dimer_hamiltonian}
\end{align}
where $\hat{n}_L = \hat{a}_L^{\dagger}\hat{a}_L$ and $\hat{n}_R = \hat{a}_R^{\dagger}\hat{a}_R$ are the number operators of the left and the right cavities, respectively. The two cavities have the same resonant frequency $\omega_0$ and Kerr non-linearity $U$, and photons can hop between the cavities at a rate $J$.

We can add a simple mechanism for incoherent driving and dissipation at the master-equation level, by using single-particle pump and loss operators. In practice, we describe the driven-dissipative dimer by a reduced density matrix $\hat{\rho}$ that evolves according to the Lindblad master equation 
\begin{equation}
    \dot{\hat{\rho}} = \doublehat{\mathcal{L}}\hat{\rho} = \doublehat{\mathcal{L}}_H\hat{\rho} + \doublehat{\mathcal{L}}_D\hat{\rho}
    \label{eq:rho_evolution}
\end{equation}
where
\begin{equation}
    \doublehat{\mathcal{L}}_H\rho = -i\left[\hat{H},\hat{\rho}\right]
    \label{eq:Lindbladian}
\end{equation}
is the Hermitian part of the evolution, while the dissipative piece reads as
\begin{align}
    \doublehat{\mathcal{L}}_D\hat{\rho} = 2 \sum_{i=L,R}
        &\Bigg\lbrace \Gamma_{i} \left( \hat{a}_i \hat{\rho} \hat{a}_i^{\dagger}
        - \frac{1}{2}\left\lbrace \hat{a}_i^{\dagger}\hat{a}_i,\hat{\rho} \right\rbrace \right) \nonumber \\
        &+ P_{i} \left( \hat{a}_i^{\dagger} \hat{\rho} \hat{a}_i
        - \frac{1}{2}\left\lbrace \hat{a}_i\hat{a}_i^{\dagger},\hat{\rho} \right\rbrace \right)
        \Bigg\rbrace
    \label{eq:Lindblad_dissipator}
\end{align}
with the constraint that $P_{i} < \Gamma_{i}  \; \forall i$, as if $\exists i : P_{i} > \Gamma_{i}$ single-particle jump operators alone are no longer sufficient to provide a correct physical description of the system.

In this form, $\Gamma_{L/R}$ are interpreted as loss rates while $P_{L/R}$ as pumping rates. It is convenient to parametrize them as
\begin{align}
    \Gamma_i &= \Gamma \pm \Delta\Gamma/2, \quad &\Delta\Gamma &= \Gamma_L - \Gamma_R \\
    P_i &= P \pm \Delta P/2, \quad &\Delta P &= P_L - P_R
\end{align}
to distinguish the case in which pump/loss rates are symmetric in the dimer, $\Delta\Gamma=\Delta P=0$ or asymmetric due to an imbalance of pump and/or losses. In fact it is known~\cite{Lebreuilly2016} that for a Bose-Hubbard lattice with uniform parameters and identical single-particle pump and loss rates, i.e.\@ $\Delta\Gamma=\Delta P=0$ the structure of the stationary state density matrix is particularly simple and reads
\begin{equation}
    \hat{\rho}_{\mathrm{ss}}=\sum_N \pi_N\Ket{N}\Bra{N}
    \nonumber
\end{equation}
where $\Ket{N}$ is a Fock state with $N$ bosons and $\pi_N\sim \left(P/\Gamma\right)^N$ up to a normalization factor. We note in the above expression that $\hat{\rho}_{\mathrm{ss}}$ is independent of any Hamiltonian parameter and only set by pump/loss ratio. This implies in particular that the stationary state occupancy $n_{\alpha} = \mathrm{Tr}\left(\rho_{\mathrm{ss}}\hat{n}_{\alpha}\right)$ is equal in the two cavities and given by
\begin{equation}
    n_L = n_R = \frac{P}{\Gamma-P}
    \label{eq:occupation_sym}
\end{equation}
which coincides with the value of an uncoupled Kerr resonator. Given these results, it is clear that any non-trivial dependence from $J/U$ has to be looked for in properties other than the stationary-state observables, as we will discuss in Sec.~\ref{sec:quantum_time_dynamics} and \ref{sec:res_gf_sym}. The above result is however no longer true in presence of a finite asymmetry in the dissipative couplings, leading to $\Delta \Gamma,\,\Delta P\neq 0$, as we will see more in detail in Sec.~\ref{sec:res_quantum_ss} and \ref{sec:res_gf_asym}.

\section{Methods}
\label{sec:methods}

The vectorized version of equation \eqref{eq:rho_evolution} is solved by exact diagonalization, yielding a bi-normalized set of left and right eigenvectors ($\bra{l_{\alpha}}$ and $\ket{r_{\alpha}}$, respectively) that satisfy
\begin{equation}
    \bra{l_{\alpha}} \hat{\mathcal{L}} = \mathcal{L}_{\alpha} \bra{l_{\alpha}}
    \qquad\text{and}\qquad
    \hat{\mathcal{L}} \ket{r_{\alpha}} = \mathcal{L}_{\alpha} \ket{r_{\alpha}}
\end{equation}
where $\hat{\mathcal{L}}$ is the matrix representation of the superoperator $\doublehat{\mathcal{L}}$. The cokernel and the kernel\footnote{The left and right eigenvectors corresponding to the special eigenvalue $\mathcal{L}_{0} = 0$.} of $\hat{\mathcal{L}}$ are, respectively, the left vacuum $\bra{I}$ and the steady-state density matrix $\ket{\rho_{\mathrm{ss}}}$.

The diagonalization problem can actually be simplified by realizing that both the Hamiltonian and the dissipator posses a global gauge symmetry, expressed by an operator functional $\doublehat{\mathcal{K}}$ that commutes with $\doublehat{\mathcal{L}}$ and that acts as $\doublehat{\mathcal{K}} \bullet = -i \left[ \hat{N}, \bullet \right]$. By exploiting this symmetry the matrix $\hat{\mathcal{L}}$ can then be written in a block-diagonal form, where each block is labeled by the eigenvalues of $\doublehat{\mathcal{K}}$.

The matrix $\hat{\mathcal{L}}$ and its eigenvectors are written in a basis of Fock states, with a cutoff $N_{\mathrm{cutoff}}$ on each particle number. We've fixed $N_{\mathrm{cutoff}} = 20$ throughout the work as a good compromise between accuracy and time and memory costs; this cutoff guarantees that the error on the displayed average steady-state occupations is equal or below $2\%$, while higher but more expensive cutoffs would not visibly change the results on the Green's functions.

\subsection{Time Dynamics}
\label{sec:time_dynamics}

Having solved the eigenproblem, we can then expand \cite{Arrigoni2018}
\begin{equation}
    \Ket{\rho(t)} 
    = e^{\mathcal{L}t}\Ket{\rho(0)}
    = \sum_{\alpha}\rho_{\alpha}(t)\ket{r_{\alpha}},
\end{equation}
where
\begin{equation}
    \rho_{\alpha}(t) \doteqdot e^{\mathcal{L}_{\alpha}t}\braket{l_{\alpha}|\rho(0)} =  e^{\mathcal{L}_{\alpha}t}\rho_{\alpha}(0).
\end{equation}
We note that the form of the Lindblad equation ensures $\mathfrak{Re}\mathcal{L}_{\alpha} \leq 0 \; \forall \alpha$, which prevents the dynamics from unbounded growth with time. Again, if we can exploit the global gauge symmetry, then it is sufficient to diagonalize just the largest diagonal block of the Lindbladian. The knowledge of the time-evolution of the density matrix can then be used to calculate the time-evolution of other observables, for example the occupations of the two cavities ($i=\{L,\,R\}$):
\begin{equation}
    n_{i}(t)
    = \Tr\Big( \hat{n}_{i} \rho(t) \Big)
    = \sum_{\alpha} \Braket{I | \hat{n}_{i} | r_{\alpha}} \rho_{\alpha}(t).
\end{equation}

\subsection{Källén-Lehmann Spectral Representation of Green's Functions}
\label{sec:ed_kl_reprsentation}

Albeit not necessary if one only wants to calculate the steady-state density matrix $\ket{\rho_{\mathrm{ss}}}$, the full knowledge of the spectrum can be used to explore the Green's functions of the system. In fact, one can obtain frequency-domain expressions for the retarded and the Keldysh components of the steady-state Green's function defined respectively as
\begin{align}
    G_{AB}^R(t) &= -i\theta(t)\Braket{\left[A(t), B(0)\right]}\\
    G_{AB}^K(t) &= -i\Braket{\left\lbrace A(t), B(0)\right\rbrace}
    \label{eq:G_AB_RK_defs}
\end{align}
where the average is taken over the stationary state and the operator $A$ is evolved with the Lindbladian of the system. Upon inserting a complete set of left and right eigenvectors of the Lindbladian and going to the frequency domain by defining $G_{AB}^{R/K}(\omega) \doteqdot \int dt\, e^{i\omega t}G_{AB}^{R/K}(\omega)$, we obtain a spectral representation of those functions:
\begin{align}
    G_{AB}^R(\omega)
    &= \sum_{\alpha}\Braket{I|A|r_{\alpha}}\Braket{l_{\alpha}|B|\rho_{\mathrm{ss}}}\frac{1}{\omega - i\mathcal{L}_{\alpha}} \nonumber \\
    &- \left(\sum_{\alpha}\Braket{I|A^{\dagger}|r_{\alpha}}\Braket{l_{\alpha}|B^{\dagger}|\rho_{\mathrm{ss}}}\frac{1}{\omega + i\mathcal{L}_{\alpha}}\right)^*
    \label{eq:G_R_KL_W}
\end{align}
\begin{align}
    G_{AB}^K(\omega)
    &= \sum_{\alpha}\Braket{I|A|r_{\alpha}}\Braket{l_{\alpha}|B|\rho_{\mathrm{ss}}}\frac{1}{\omega - i\mathcal{L}_{\alpha}} \nonumber \\
    &- \sum_{\alpha}\Braket{I|B|r_{\alpha}}\Braket{l_{\alpha}|A|\rho_{\mathrm{ss}}}\frac{1}{\omega + i\mathcal{L}_{\alpha}} \nonumber \\
    &+ \left(\sum_{\alpha}\Braket{I|A^{\dagger}|r_{\alpha}}\Braket{l_{\alpha}|B^{\dagger}|\rho_{\mathrm{ss}}}\frac{1}{\omega + i\mathcal{L}_{\alpha}}\right)^* \nonumber \\
    &- \left(\sum_{\alpha}\Braket{I|B^{\dagger}|r_{\alpha}}\Braket{l_{\alpha}|A^{\dagger}|\rho_{\mathrm{ss}}}\frac{1}{\omega - i\mathcal{L}_{\alpha}}\right)^*
    \label{eq:G_K_KL_W}
\end{align}
We see that the Green's functions of an open Markovian quantum system can be generically written as sum of simple poles at complex frequencies given by the eigenvalues of the Lindbladian and with weights, in general complex, given by the transition matrix elements between the stationary state and some excited state of the system \cite{Arrigoni2018,Scarlatella2018}.

From the practical point of view, if one focuses on the single-particle Green's functions, the calculation can be further simplified via the block-diagonal structure of the Lindbladian outlined above. In fact, since the calculation of the single-particle Green's functions involves states that differ at most by one particle from the stationary state, it turns out that the full knowledge of the spectrum is not necessary; it is sufficient to diagonalize just the 3 largest blocks of the diagonal-block structure. Assuming that the diagonlization scales as the cube of the matrix linear dimension, this yielded a theoretical $10^4$ speedup of the diagonalization with the 20-bosons cutoff we have used in both cavities, as well as a $99.7\%$ reduction of the memory required to store the results.


\section{Review of Semiclassical Dynamics and Self-Trapping Transition}
\label{sec:semiclassical_dynamics_recap}

In order to have a reference point for the analysis of our results we can start by recalling the predictions of a semiclassical treatment of the quantum dynamics for the BHD~\cite{Smerzi1997,Sarchi2008}. This is obtained by writing the exact equations of motion for the cavity field operators $\hat{a}_{L/R}$ and by closing them by taking $\hat{a}_{L/R} \equiv \alpha_{L/R}$, where $\alpha_{L/R}$ are $c$-numbers. It is important to remark that this approach, which assumes a coherent state of bosons, works for large photons number, while in the quantum treatment we are typically interested in a few-photons treatment. The resulting equations of motion read
\begin{align}
    \dot{\alpha}_{L} 
    &= -i \Big[ (\omega_0-U) + 2U\vert\alpha_L\vert^2  \Big] \alpha_L - iJ\alpha_R - \Gamma_{L}^{\mathrm{eff}}\alpha_L \nonumber \\
    \dot{\alpha}_{R} 
    &= -i \Big[ (\omega_0-U) + 2U\vert\alpha_R\vert^2  \Big] \alpha_R - iJ\alpha_L - \Gamma_{R}^{\mathrm{eff}}\alpha_R \nonumber
\end{align}
where $\Gamma_{L/R}^{\mathrm{eff}} = \Gamma_{L/R} - P_{L/R}$ are the \emph{effective} loss rates, which for single-particle losses must always be positive.

\begin{figure}
    \centering
    \includegraphics[scale=0.49]{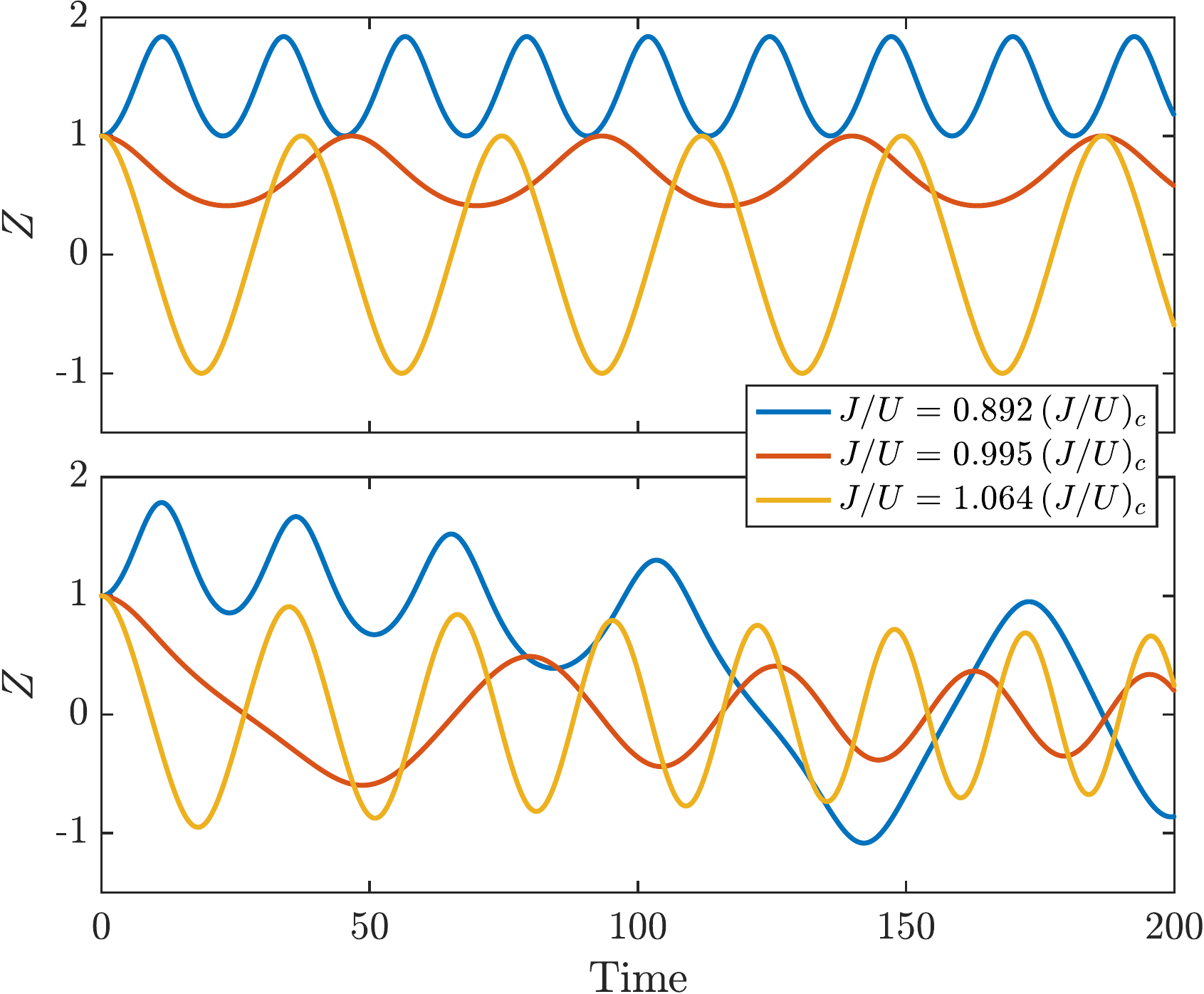}
    \caption{Evolution of the occupation imbalance for $N_0=3$, $Z_0=1$, $U=0.1$ and $\Delta\omega=0$. Different colors correspond to different values of $J/U$ around the critical value, predicted via Eq.~\eqref{eq:J_over_U_semi_critical}. The lines in the top panel are for a closed system, while the ones of the same color in the bottom panel are for an open system with $\Gamma^{\mathrm{eff}}_L = \Gamma^{\mathrm{eff}}_R = 4 \times 10^{-4}$; they are obtained respectively by numerically solving the full system of equations discussed in Sec.~\ref{sec:semiclassical_dynamics_recap} (see also Eq.~\eqref{eq:BH_dimer_semiclassical_equations}-\eqref{eq:BH_dimer_semiclassical_equations_hamiltonian} in Appendix~\ref{app:semiclassical_dynamics}).
    }
    \label{fig:Z_vs_time_semiclassical}
\end{figure}
As discussed in more detail in Appendix~\ref{app:semiclassical_dynamics}, it's possible to write semiclassical equations for the total number of photons $N = n_L + n_R$ and for the occupation imbalance between the two cavities $Z = n_L - n_R$, with $n_{L/R}=\vert\alpha_{L/R}\vert^2$.

In the closed-system case, corresponding to $\Gamma_{L/R}^{\mathrm{eff}}=0$, number and energy conservation yield simplified analytical results for the imbalance $Z$, predicting a transition from a regime in which $Z$ oscillates above the initial condition $Z_0$ to a regime in which it oscillates around $0$ (solid lines in Fig.~\ref{fig:Z_vs_time_semiclassical}) as one increases the value of $J/U$ above the critical coupling
\begin{equation}
    \left(\frac{J}{U}\right)_{\mathrm{c}} = N_0\left( \frac{\sqrt{1-(Z_0/N_0)^2} + 1}{2} \right).
    \label{eq:J_over_U_semi_critical}
\end{equation}
which depends on the initial total number of photons $N_0$ and imbalance $Z_0$. This phase transition can be seen as a divergence of the oscillation period (Fig.~\ref{fig:semiclassical_T_divergence}) or as a sharp decay to zero of the time-averaged imbalance $\braket{Z}_T = \frac{1}{T}\int_{t_0}^{t_0+T} dt\, Z(t)$ (Fig.~\ref{fig:divergence_semi}, bottom panel) (see Appendix~\ref{app:semiclassical_dynamics}).

\begin{figure}
    \centering
    \includegraphics[scale=0.49]{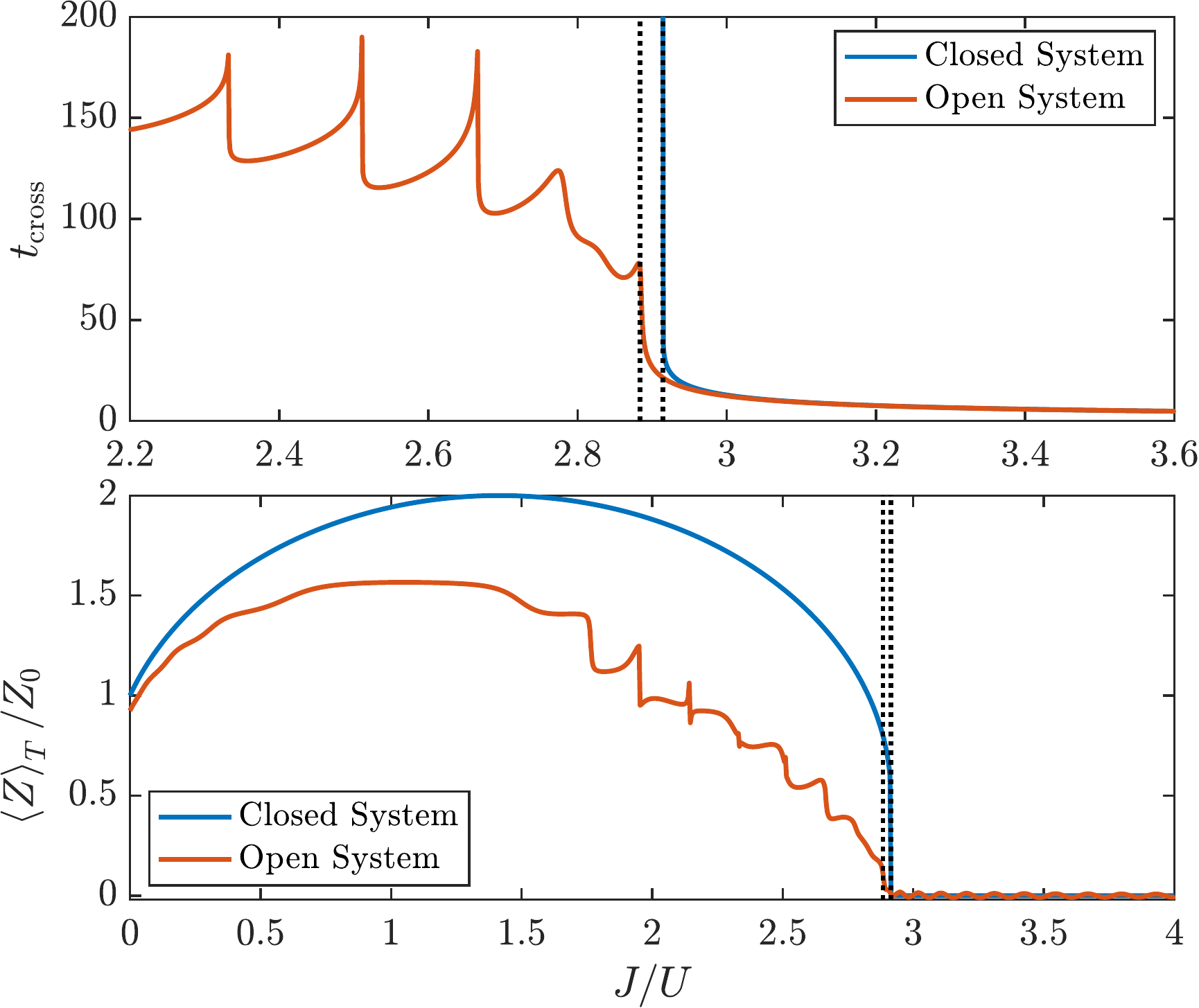}
    \caption{(Top)
    Time $t_{\mathrm{cross}}$ at which $Z(t)$ crosses the value $Z=0$, as a function of $J/U$.
    (Bottom) Semiclassical time-averaged imbalance. While the closed system has an analytical expression, the open case requires to solve the full dynamics \eqref{eq:BH_dimer_semiclassical_equations} and to choose an upper time limit in the integration; in this plot, we integrate up to $t = 200$.
    In both panels, the critical value of $J/U$ predicted via \eqref{eq:J_over_U_semi_critical} (for the closed system) and as a numerical estimate (for the open system) is shown as a vertical dotted line.}
    \label{fig:divergence_semi}
\end{figure}

The open system case is not analytically solvable, but the numerical solution of the equations for the total number of photons and for the cavity occupation imbalance shows that the closed-system picture is preserved for low enough values of the loss coefficients, with the difference that even oscillations around a value that is different from zero at initial times will eventually transition at long enough times to an oscillation regime around zero during the dynamical evolution (Fig.~\ref{fig:Z_vs_time_semiclassical}, bottom panel).

We can define the time at which this dynamical transition happens to be some $t_{\mathrm{cross}}$ for which the imbalance $Z(t)$ crosses the value $Z=0$ for the first time. If we plot this time as a function of $J/U$, see top panel of Fig.~\ref{fig:divergence_semi}, we expect that for the closed system this time is divergent for values of $J/U$ below the critical value; for the open system, however, this time assumes finite values even below the critical point and the critical point itself is at a slightly lower value than its closed-system counterpart ($(J/U)_c=2.88$ vs.\@ $(J/U)_c=2.91$). The peak structure visible below $(J/U)_c$ for the open system is due to the commensurability between the period of the imbalance oscillations, that is a function of $J/U$ itself, and $t_{\mathrm{cross}}$.

Albeit holding in the limit of large photon number only, these semiclassical results provide a useful hint for the quantities to look at in the quantum case, as well as a point of comparison that highlights the intrinsic differences between the two types of analyses.

\section{Results: Dissipative Quantum Dynamics}
\label{sec:quantum_time_dynamics}

We now move on to discuss the full dissipative quantum dynamics of the BHD introduced in Sec.~\ref{sec:model}. We focus in particular on the occupation imbalance $Z(t) = n_L(t) - n_R(t)$ between the two cavities, which in the semiclassical limit shows a clear change of behavior as a function of the parameters.

In the following we set $\omega_0=1$, $U=0.1$ and consider a situation of symmetric pump and loss rates, $\Delta P = \Delta\Gamma = 0$, so that by construction the imbalance is zero at long times. We set the effective losses $\Gamma_{L/R}^{\mathrm{eff}} = \Gamma_{L/R} - P_{L/R} = 1 \times 10^{-4}$ and the pump $P_{L/R} = 2 \times 10^{-4}$, such that the identical occupation in the two cavities is $n_L = n_R = 2$ (see Eq.~\eqref{eq:occupation_sym}, independently on $J/U$.

\begin{figure}
    \centering
    \includegraphics[scale=0.49]{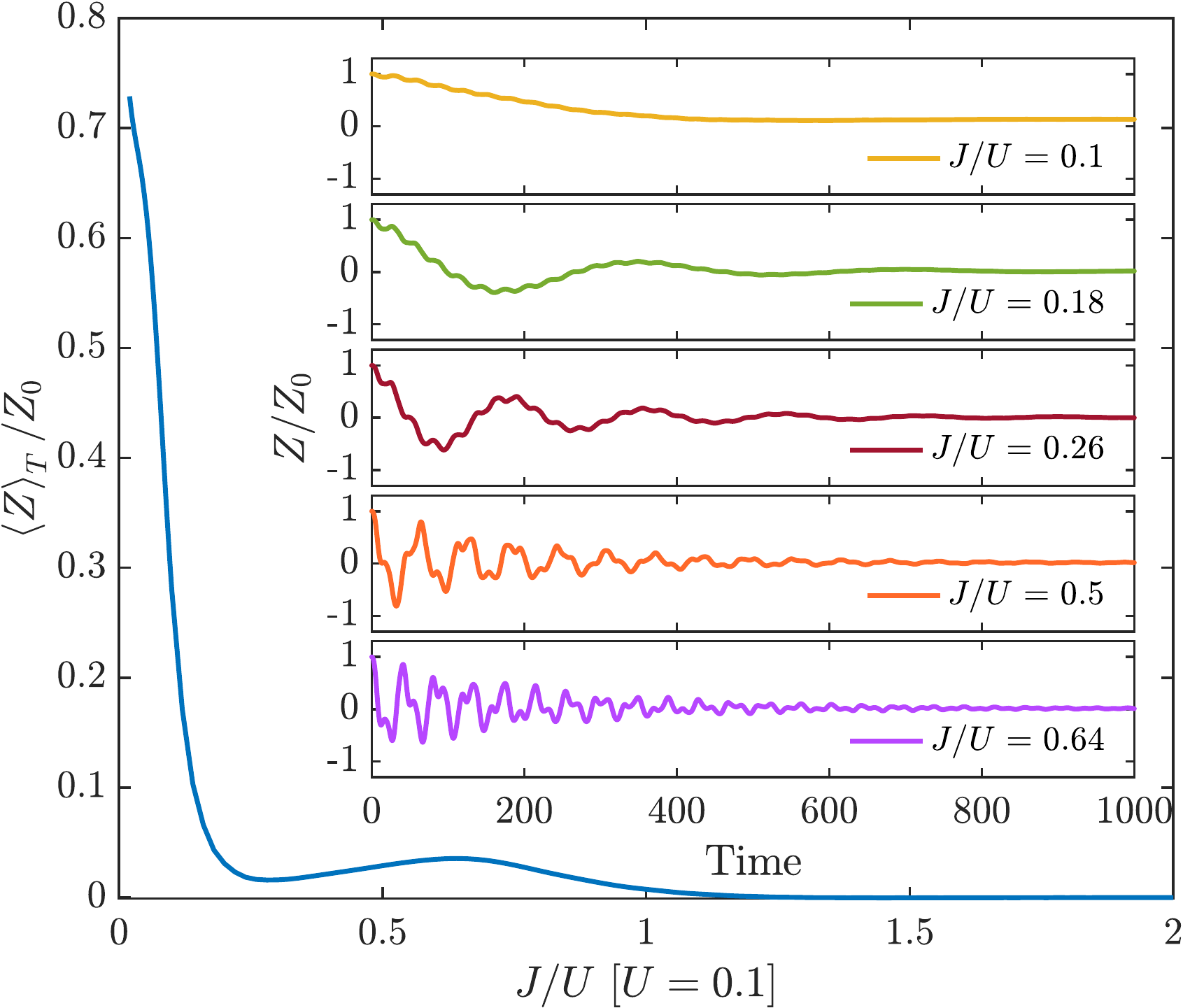}
    \caption{
    (Inset) Imbalance $Z(t) = n_L(t)-n_R(t)$ for different values of $J/U$ [$U=0.1$], starting from a state $\Ket{3,1}$ at $t=0$ ($Z_0 = 2$). The cavities have a base frequency $\omega_0=1.0$. The effective loss is $\Gamma_L^{\mathrm{eff}} = \Gamma_R^{\mathrm{eff}} = 1 \times 10^{-4}$ and the pumping rate realizes a steady-state occupation equal to $2$ in both cavities, so that $Z_{\mathrm{ss}} = 0$ by construction. The semiclassical, non-dissipative critical value of $J/U$ for this particular configuration is $(J/U)_c \approx 3.73$.
    The time-averaged occupation imbalance computed over the time interval $[0,\,1000]$ is shown in the main panel.}
    \label{fig:multiJ_Imb}
\end{figure}

We start discussing the imbalance dynamics as a function from $J/U$, at a fixed initial condition which we take to be a Fock state $\Ket{3,1}$, corresponding to an initial imbalance $Z_0 = 2$ and an initial number of photons $N_0 = 4$. At the semiclassical level, see Eq.~\eqref{eq:J_over_U_semi_critical}, this would correspond to a critical coupling $(J/U)_c=3.73$ for the self-trapping transition.

In the inset of Fig.~\ref{fig:multiJ_Imb} we plot the time-dependent imbalance $Z(t)$ for different values of $J/U$. We find a clear crossover as the hopping is increased, from a pure exponential decay to zero at small $J/U=0.1$, to an underdamped decay with fast oscillations superimposed at $J/U=0.26$ which evolves further into strongly anharmonic oscillations at large values of the hopping, whose frequency grows with $J/U$. We can interpret this behavior as a signature of the self-trapping transition in the dissipative quantum dynamics. In the small hopping regime each site of the dimer evolves almost independently and the imbalance goes to zero, while for larger values of the hopping there is a substantial transfer of photons across the dimer, resulting in coherent Rabi-like oscillations, before the imbalance reaches the stationary state.

The $J/U$ dependence can also be studied from the point of view of the time-averaged occupation imbalance $\Braket{Z}_T$. In contrast to the semiclassical case (Fig.~\ref{fig:divergence_semi}), where one expects a sharp transition\footnote{In the open case, the extent of the jump discontinuity in $\partial_{J/U}\Braket{Z}_T$ depends on the upper limit of the integration time.} between $\Braket{Z}_T \neq 0$ and $\Braket{Z}_T = 0$, in the quantum case we have a smooth crossover between the two regimes. The average imbalance drops quickly with $J/U$ due to the development of damped Rabi oscillations, reaching a minimum around $J/U\simeq 0.25$. Quite interestingly, though, we find the appearance of a region in which the imbalance actually \emph{increases} as a function of $J/U$ before completely dropping to $0$ at higher values of $J/U$. We note that, with respect to the semiclassical case, the localized (self-trapped) phase with $\Braket{Z}_T \neq 0$ is strongly suppressed and that already for $J/U\simeq 1.25$ the average imbalance is zero. This is consistent with the expectation that quantum fluctuations, included in the exact solution and not properly treated in the semiclassical approach, tend to reduce the broken symmetry phase.
 
\begin{figure}
    \centering
    \includegraphics[scale=0.49]{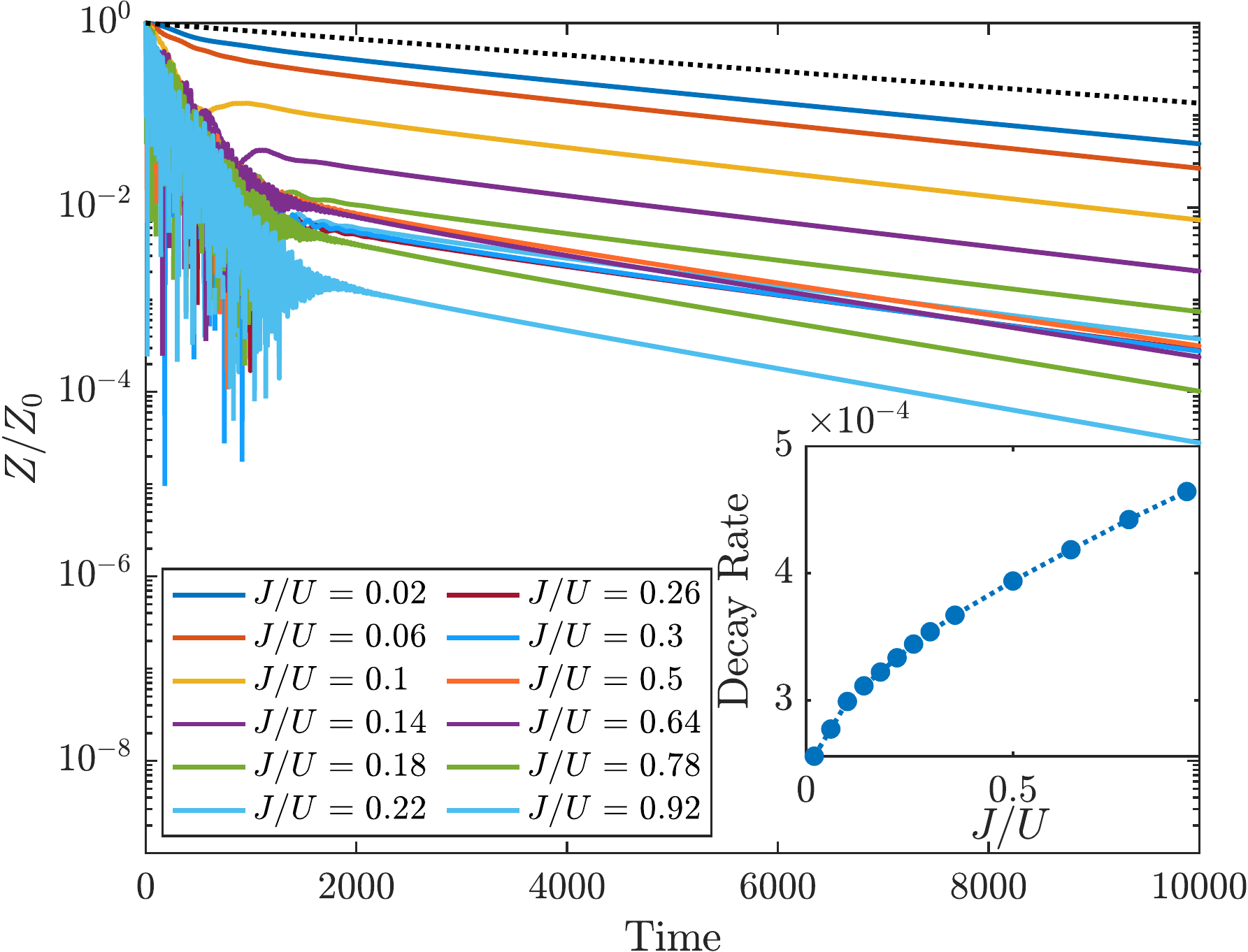}
    \caption{
    Evolution of the imbalance $Z(t)$ for the same settings of Fig.~\ref{fig:multiJ_Imb}, shown at longer times and at log scale. The black, dotted line is obtained analytically at $J/U=0$; it corresponds to an exponential decay at a rate $2\Gamma^{\mathrm{eff}}$. At long times, $\mathrm{ln}\left(Z(t)/Z_0\right)$ fits a straight line; the inset shows the corresponding decay rate as a function of $J/U$.}
    \label{fig:multiJ_Imb_log}
\end{figure}

We now discuss the dynamics on longer time scales, where we expect the small dissipative couplings to dominate over the Hamiltonian parameters. To this extent in Fig.~\ref{fig:multiJ_Imb_log} we plot the time-dependent imbalance over a broad range of time scales and for different values of $J/U$. We see a clear separation of dynamical regimes, from a short-time one - strongly dependent on $J/U$, as we discussed above - to a longer-time one where the imbalance exponentially decays to zero. While naively one could have expected the decay rate to be set only by the dissipative couplings we see in the inset of Fig.~\ref{fig:multiJ_Imb_log} that instead it shows a monotonic increase with $J/U$. 
 
Finally, we consider the dependence of the time-dependent imbalance $Z(t)$ from the initial condition. To this extent we fix as initial density matrix a pure Fock state $\rho_0=\Ket{n_{0L},n_{0R}}\Bra{n_{0L},n_{0R}}$, corresponding to an initial imbalance $Z_0=n_{0L}-n_{0R}$ and initial photon number $N_0=n_{0L}+n_{0R}$, and change the values of $n_{0L},n_{0R}$. At the semiclassical level, as we see in Eq.~\eqref{eq:J_over_U_semi_critical}, there is a critical value of $J/U$ for any $N_0,Z_0$. In order to highlight the difference between the exact quantum dynamics and the semiclassical evolution we fix the value of the hopping to interaction ratio $J/U$ to be always below $(J/U)_c(N_0,Z_0)$, such that at the semiclassical level the system should be localized (self-trapped) at short times for all the chosen initial conditions (see Eq.~\eqref{eq:J_over_U_semi_critical}) and delocalized at longer times (see Fig.~\ref{fig:Z_vs_time_semiclassical}). 

\begin{figure}[t]
    \centering
    \includegraphics[scale=0.49]{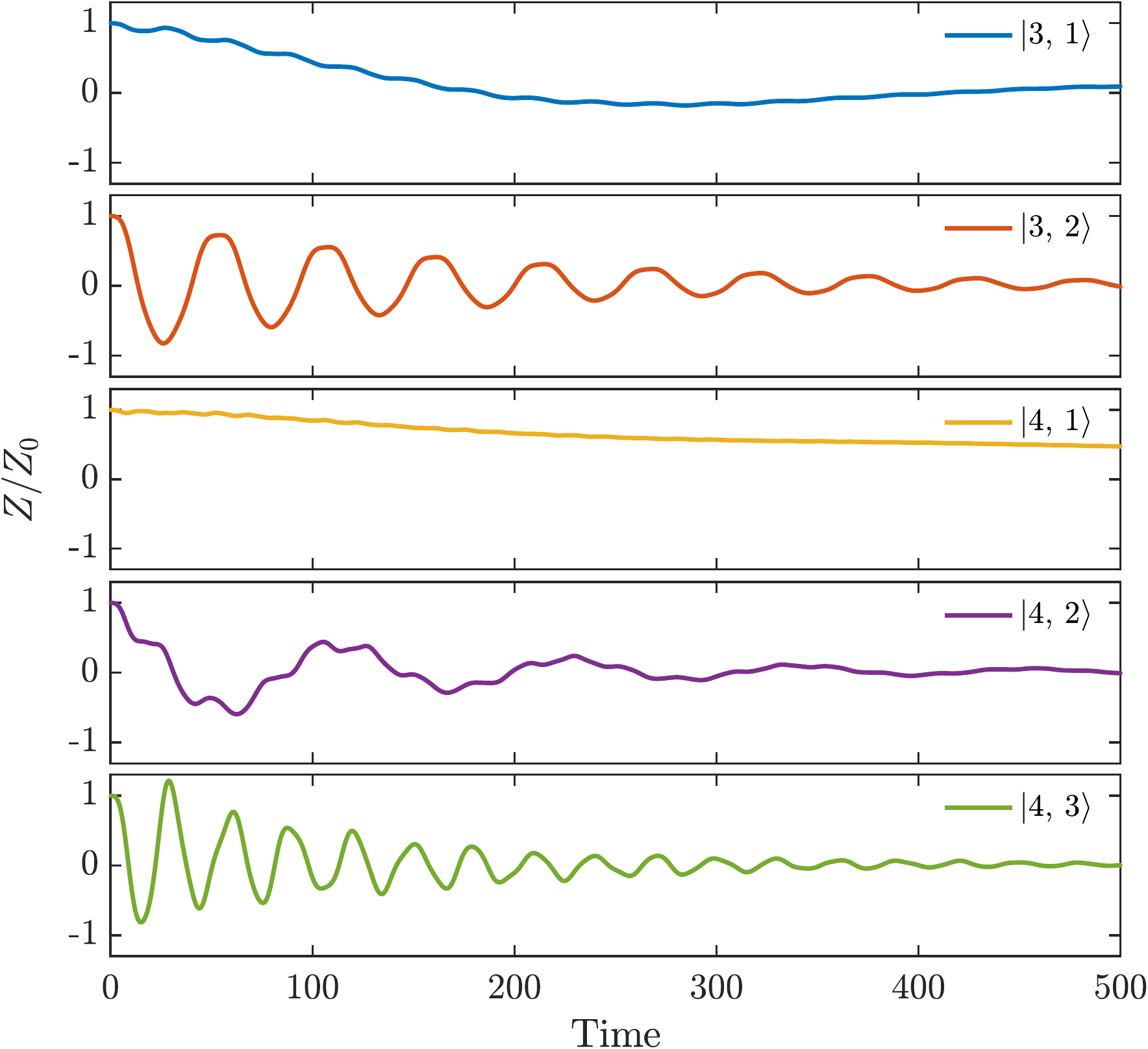}
    \caption{Imbalance $Z(t) = n_L(t)-n_R(t)$ for $(J/U)=0.04(J/U)_c$ [$U=0.1$], starting from different $\Ket{n_{0L},n_{0R}}$ number states at $t=0$. The cavities have a base frequency $\omega_0=1.0$. The effective loss is $\Gamma_L^{\mathrm{eff}} = \Gamma_R^{\mathrm{eff}} = 1 \times 10^{-4}$ and the pumping rate realizes a steady-state occupation equal to $2$ in both cavities, so that $Z_{\mathrm{ss}} = 0$ by construction. The quantity $(J/U)_c$ refers to the semiclassical non-dissipative value in \eqref{eq:J_over_U_semi_critical}.}
    \label{fig:multiKet_Imb}
\end{figure}

We plot in Fig.~\ref{fig:multiKet_Imb} the quantum dynamics of the imbalance for different initial conditions. We see that, quite at the opposite of what expected from the semiclassical analysis, the evolution of $Z(t)$ has a strong dependence on the initial state in which the system is prepared. In particular we find both regimes of slow decay to zero of the imbalance (see for example the initial conditions corresponding to $\Ket{3,1}$ or $\Ket{4,1}$), indicating localized/self-trapped behavior, as well as regimes of coherent Rabi-like oscillations of the imbalance (see for example the initial conditions corresponding to $\Ket{3,2}$ or $\Ket{4,3}$) that we can interpret as signatures of delocalization. This is consistent with the observation made earlier (see Fig.~\ref{fig:multiJ_Imb}) that quantum fluctuations renormalize the critical coupling and favor the delocalized regime. We conclude therefore that, as in the semiclassical case, the self-trapping crossover can be accessed by changing the initial condition, however we do not explore here the precise dependence of $\left(J/U\right)_c$ from the initial state and whether it can be encoded in a simple expression depending only on $N_0$ and $Z_0$ as in Eq.~\eqref{eq:J_over_U_semi_critical}.

\begin{figure}[t]
    \centering
    \includegraphics[scale=0.49]{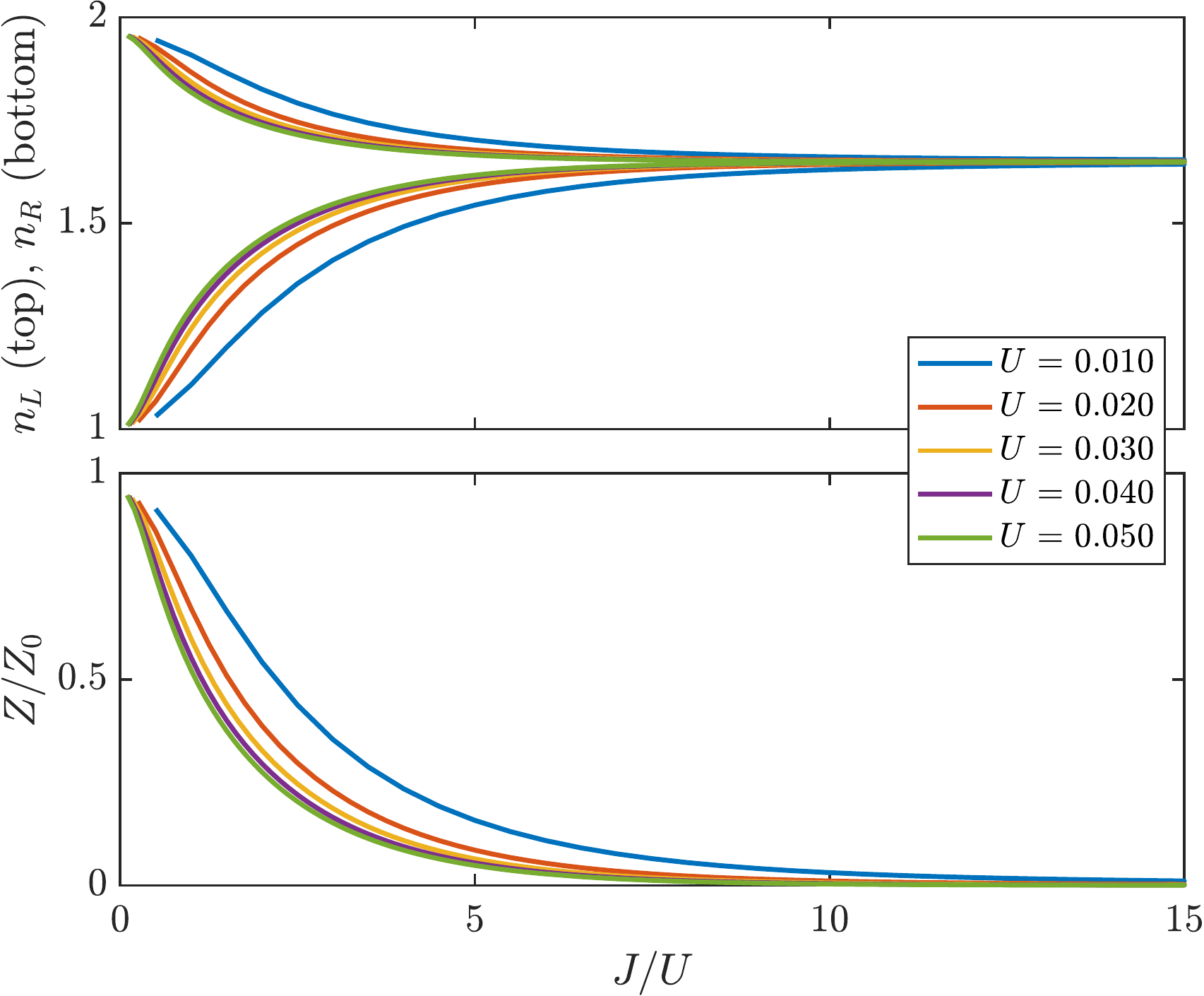}
    \caption{(Top) Steady-state cavity occupations as a function of $J/U$ for a dimer with loss coefficients $(\Gamma_L,\,\Gamma_R) = (6\times10^{-2},\,2\times10^{-2})$ and pump coefficients $(P_L,\,P_R) = (4\times10^{-2},\,1\times10^{-2})$ described in the main text, at different values of $U$. The cavities have a base frequency $\omega_0=1.0$. The top curves are the occupations of the left cavity, while the bottom ones are the occupations of the right cavity.
    (Bottom) Steady-state imbalance $Z = n_L - n_R$ corresponding to the occupations in the top panel.
    }
    \label{fig:2Cav_SS_midlow}
\end{figure}

\section{Results: Quantum Steady State for finite Pump/Loss Asymmetry}
\label{sec:res_quantum_ss}

In the previous section we have considered the case of a BHD with symmetric pump and loss rates, resulting in a trivial stationary state with zero imbalance for any value of $J/U$, but with a rich nonequilibrium dynamics. 

As we discussed in Sec.~\ref{sec:model}, in presence of a finite pump/loss asymmetry among the two cavities the stationary state becomes more interesting. We can therefore look for signatures of a delocalization crossover, analogous to what we have shown in Fig.~\ref{fig:multiJ_Imb}, directly in observables such as the steady-state occupation or imbalance.

As an example, we consider two cavities with loss coefficients $(\Gamma_L,\,\Gamma_R) = (6\times10^{-2},\,2\times10^{-2})$ and pump coefficients $(P_L,\,P_R) = (4\times10^{-2},\,1\times10^{-2})$, that thus realize steady-state occupations $(n_{0L},n_{0R}) = (2,1)$ in the uncoupled limit $J=0$ (see Eq.~\eqref{eq:occupation_sym}). In Fig.~\ref{fig:2Cav_SS_midlow} we plot the dependence of the two cavity occupations (top panel) and imbalance (bottom panel) from the hopping to interaction ratio $J/U$, for different values of $U$ (keeping $\omega_0=1$ as unit). We see in the top panel that as $J/U$ is increased the two occupations both converge towards a common value, which is essentially independent from $U$. The large-$J/U$ limit of the occupations can be obtained analytically by considering the limit $U=0$ and results in a weighted average of the two uncoupled occupations (see Eq.~\eqref{eq:weighted_mean_occupations}). 

As a consequence of the two occupations becoming equal at large $J/U$ we see in the bottom panel that the steady-state imbalance between the two cavities reduces and approaches zero for large enough $J/U$, a signature of delocalization. We note that increasing $U$ pushes the crossover $J/U$ scale for delocalization to lower values and we expect for $U=0.1$ to obtain a behavior comparable with what obtained from the dynamics (see Fig.~\ref{fig:multiJ_Imb}).

\section{Results: Green's Functions}
\label{sec:res_gf}

A way to get some insights on the system even when the steady-state observables do not depend neither on $J$ nor on $U$, as in the case of symmetric pump and losses, is to look instead at the single-particle Green's functions. Either by seeing them as the resolvent of the Lindbladian or as response functions that link different states and thus participate in the calculation of transport quantities like the optical transmission, the Green's functions are sensitive to the details of the Lindbladian spectrum, and not only to the zero mode (stationary state), as it appears clearly from the Källén-Lehmann representation discussed in Sec.~\ref{sec:ed_kl_reprsentation}.

In this section we present our results for the Green's function of the BHD, that we obtained from the exact diagonalization of the Lindbladian as discussed in Sec.~\ref{sec:methods}. Specifically we consider the single-particle Green's functions, obtained from Eq.~\eqref{eq:G_K_KL_W} with the choice $A = a_i$ and $B = a_j^\dagger$ with $i,j=L/R$, and in particular the spectral function $\mathcal{A}_{ij}(\omega)$ and the cavity correlation function $\mathcal{C}_{ij}(\omega)$, defined as
\begin{equation}
    \mathcal{A}_{ij}(\omega) \doteqdot - \frac{1}{\pi} \Imag G^R_{ij}(\omega),
    \quad\ 
    \mathcal{C}_{ij}(\omega) \doteqdot -\frac{1}{2\pi i} G^K_{ij}(\omega)
    \label{eq:spectralA_L__cavcorfunC_L}
\end{equation}
with $i,j=L/R$. The diagonal components (for $i=j$) contain information on the local (on-site) spectrum and occupations of the bosonic mode and satisfy the sum rules
\begin{align}
    \int_{-\infty}^{+\infty} d\omega \mathcal{A}_{i}(\omega) 
    &= 1 \\
    \int_{-\infty}^{+\infty} d\omega\, \mathcal{C}_{i}(\omega)
    &= 2n_i + 1
    \label{eq:n_via_cavcorfun}
\end{align}
where $n_i$ is the stationary state occupation. The off-diagonal components contain instead information on the delocalized modes across the dimer. In particular the correlation function $C_{LR}(\omega)$ has real and imaginary parts which satisfy the sum-rules
\begin{align}
    J\int_{-\infty}^{+\infty} d\omega\, \Real \mathcal{C}_{LR}(\omega)
    = \braket{\hat{T}}
    \label{eq:sum_rule_ReC} \\
    J\int_{-\infty}^{+\infty} d\omega\, \Imag \mathcal{C}_{LR}(\omega)
    = \braket{\hat{I}}
    \label{eq:sum_rule_ImC}
\end{align}
where $\braket{\hat{T}} = J\braket{\hat{a}^{\dagger}_L\hat{a}_R+\hat{a}^{\dagger}_R\hat{a}_L}$ is the average kinetic energy in the stationary state while $\braket{\hat{I}} = -iJ \braket{\hat{a}^{\dagger}_R\hat{a}_L-\hat{a}^{\dagger}_L\hat{a}_R}$ is the average current flowing from $L$ to $R$ (see Appendix~\ref{app:sumrules}). We now presents our results for these Green's functions, starting from the pump/loss symmetric case and then discussing the role of a finite pump/loss asymmetry.

\begin{figure}[!htbp]
    \centering
    \includegraphics[scale=0.49]{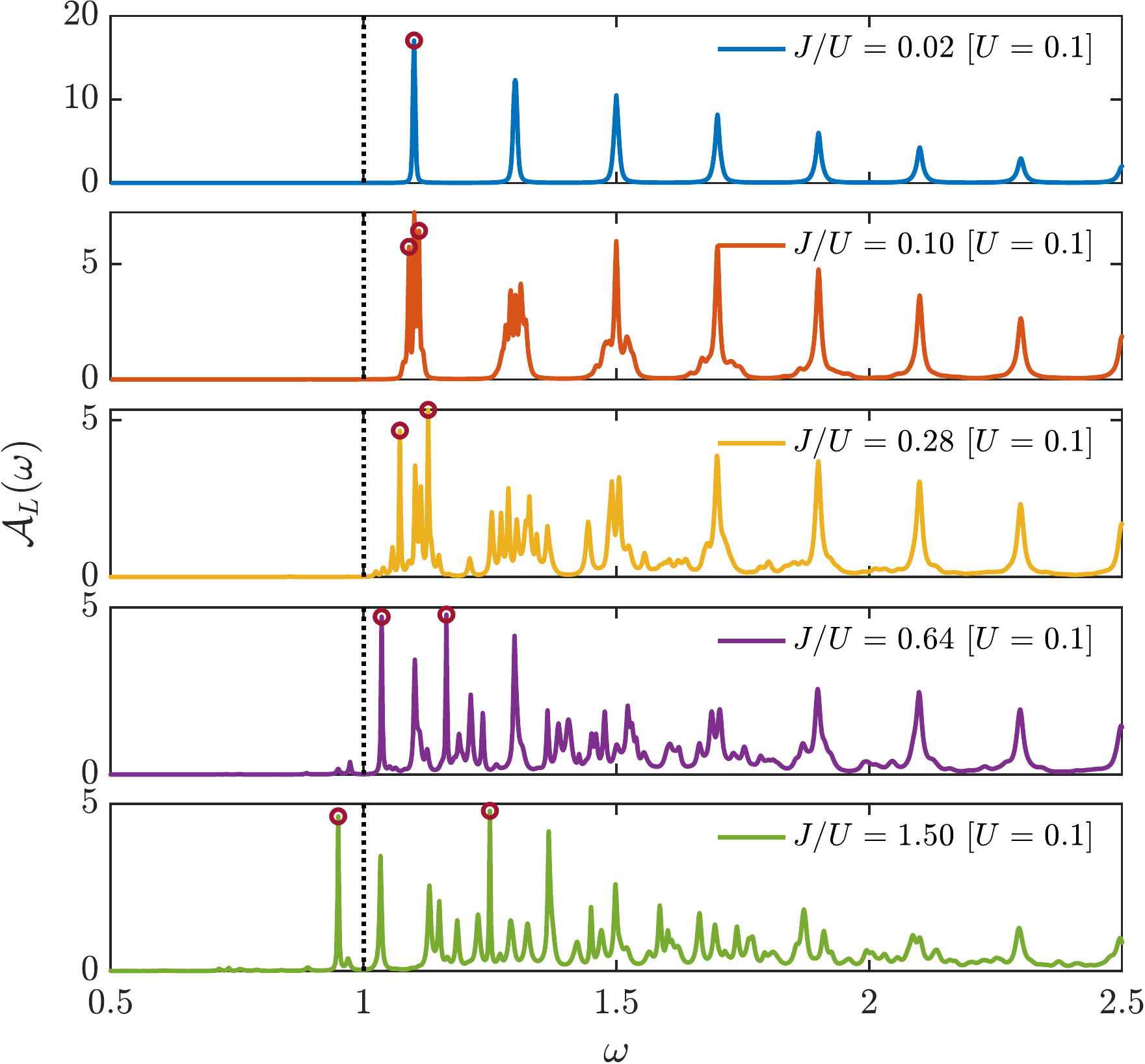}\\\medskip
    \includegraphics[scale=0.49]{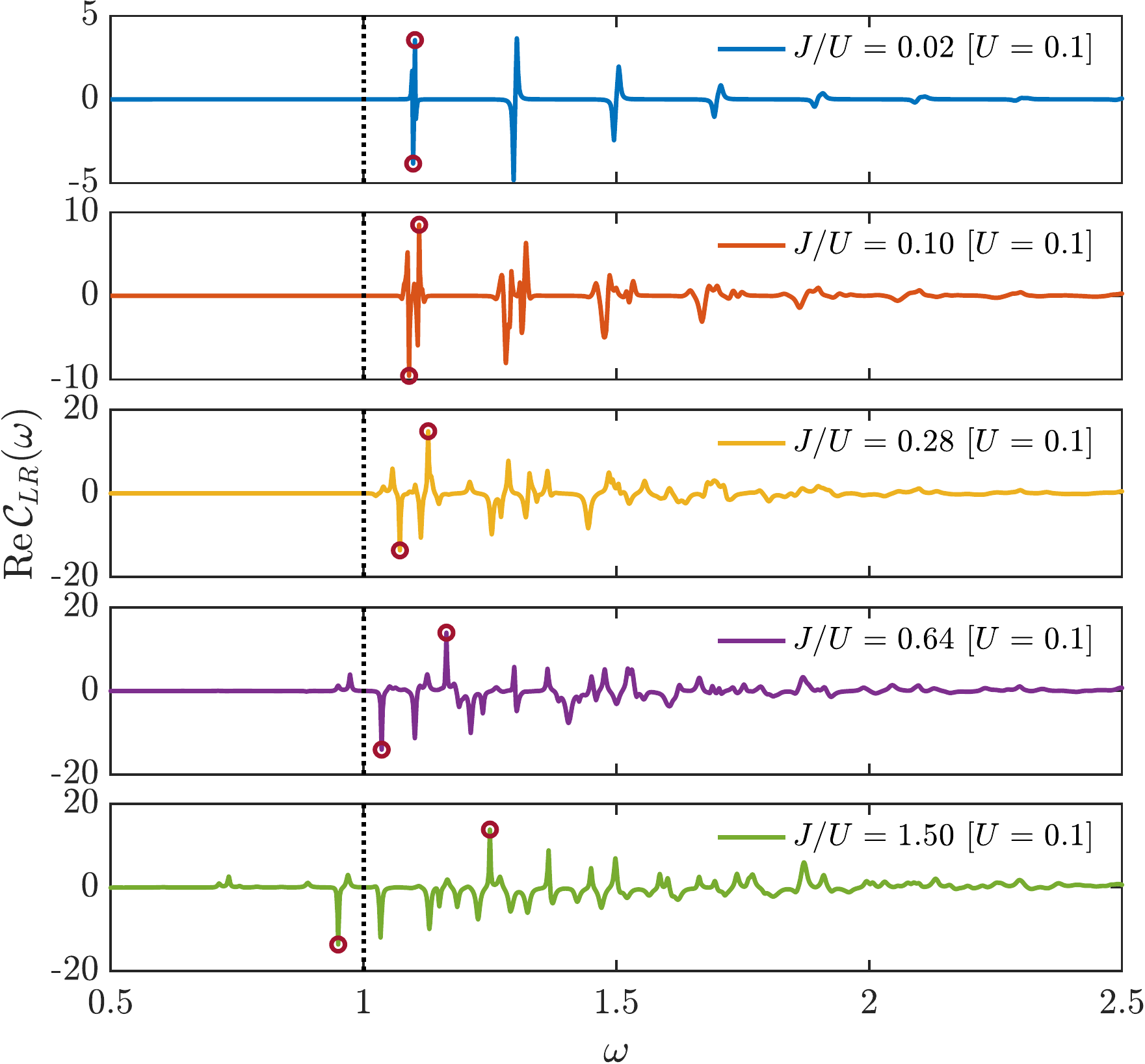}
    \caption{(Top) Spectral function $\mathcal{A}_L(\omega)$ for different values of $J/U$ [$U=0.1$]. The cavities have a base frequency $\omega_0=1.0$ (vertical dotted line), while the effective loss is $\Gamma_L^{\mathrm{eff}} = \Gamma_R^{\mathrm{eff}} = 1 \times 10^{-4}$ and the pumping rate realizes a steady-state occupation equal to $2$ in both cavities. The circled peaks mark the bonding/anti-bonding states resulting from the splitting of the first excited state at $\omega_0+U$ for decoupled cavities.
    (Botttom) Real part of the off-diagonal cavity correlation function $\mathcal{C}_{LR}(\omega)$. This function is negative (positive) for the bonding (anti-bonding) states marked by circles and discussed in the top panel..}
    \label{fig:multiGF_Imb}
\end{figure}


\subsection{Symmetric Pump and Losses}
\label{sec:res_gf_sym}

We start considering the case of symmetric pump and loss rates, $\Delta\Gamma=\Delta P=0$. As a result the system is completely symmetric upon reflection ($L\leftrightarrow R$) and as such the diagonal spectral functions in Eq.~\eqref{eq:spectralA_L__cavcorfunC_L} do not depend on the index $i=L/R$. As an example, in the top panel of Fig.~\ref{fig:multiGF_Imb} we plot the spectral function of the left cavity for different values of $J/U$. 

At low $J/U$ the spectral function resembles much the one of a single driven-dissipative Kerr resonator, with a characteristic sequence of peaks located at frequencies given by the energy difference between states with $n+1$ and $n$ photons, $\Delta_n=E_{n+1}-E_n=\omega_0+U+2Un$, where $E_n=\omega_0 n+Un^2$ is the energy of the Kerr resonator with $n$ photons (see the Hamiltonian in Eq.~\eqref{eq:dimer_hamiltonian}). These peaks, which start at $\omega_0+U$ and are equally spaced by $2U$, would be infinitely sharp in the closed system while are broadened by the dissipative processes by an amount roughly given by $\Gamma_L^{\mathrm{eff}}$ (it would perfectly match this value in the non-interacting, decoupled case $J=U=0$, see Appendix~\ref{app:greens_U0}.)

As $J/U$ is increased we see that the first effect is the creation of sub-peaks within each resonance, particularly in the low frequency ones, with the center of mass of each \emph{band} remaining roughly located at the isolated Kerr excitation energies. Upon increasing further $J/U$ we see how different bands start to merge in a continuum and for $J/U=0.64$ a new features arises, namely a finite spectral weight appears below the resonator frequency $\omega_0=1$, which becomes a sharp peak for large values of $J/U$ (e.g.\@ $J/U=1.50$). This peak corresponds to a delocalized photonic excitation as one can realize by looking at the spectral function in the opposite limit of $U=0$ (see Appendix~\ref{app:greens_U0}), which has two poles at frequencies roughly $\omega_{\pm} \simeq \omega_0 \pm J$ since in this regime the dissipative couplings are very small. 

It is interesting to connect these spectral features to the behavior of the time-dependent and of the time-averaged imbalance shown in Fig.~\ref{fig:multiJ_Imb} for similar values of $J/U$. For small values of the hopping the imbalance is different from zero at short and intermediate times, i.e.\@ photons remain localized in one of the two cavities and the spectral function resembles the one of an isolated Kerr resonator. Upon increasing $J/U$ photons start to hop coherently within the dimer: the imbalance shows short-time Rabi oscillations with a period controlled by $J/U$ and its time-average vanishes, while spectrally this translates in the emergence of two peaks above and below the bare resonator frequency.

\begin{figure}[b]
    \centering
    \includegraphics[scale=0.45]{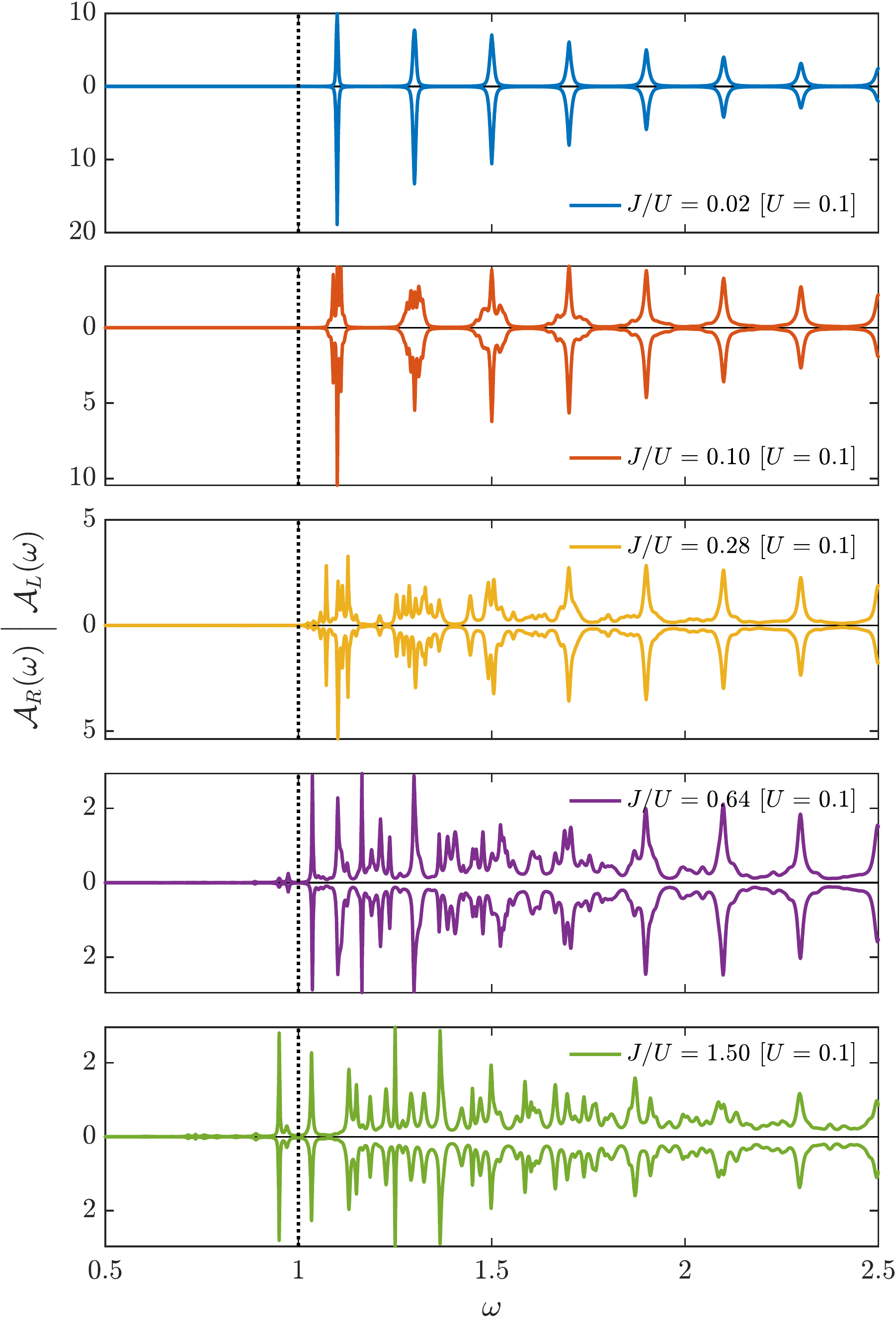}
    \caption{Spectral functions $\mathcal{A}_L(\omega)$ (top) and $\mathcal{A}_R(\omega)$ (bottom) for different values of $J/U$ [$U=0.1$]. The cavities have a base frequency $\omega_0=1.0$ (vertical dotted line), while the effective losses are $\Gamma^{\mathrm{eff}}_L = 2.5 \times 10^{-4}$ and $\Gamma^{\mathrm{eff}}_R = 1 \times 10^{-4}$ and the pumping rates realizes uncoupled steady-state occupations equal to $\sim 3.3$ in the left cavity and $2$ in the right cavity.}
    \label{fig:multiGF_DP_Imb}
\end{figure}

In the bottom panel of Fig.~\ref{fig:multiGF_Imb} we plot the real-part of the off-diagonal correlation function, for different values of $J/U$ and $\Delta\Gamma=\Delta P=0$. We note that quite interestingly the imaginary part of this Green's function vanishes in this regime, a point onto which we will come back in the next section. At small values of the hopping the real-part $\mathcal{C}_{LR}(\omega)$ is essentially zero, the cavities are almost decoupled, except at frequencies corresponding to the eigenmodes of the (interacting) single cavity (see top panel at the same value of $J/U$), where an anti-resonance like contribution emerges. Upon increasing $J/U$, as we discussed for the spectral function, further peaks appear which start merging and shifting towards lower frequencies. We note that the structure of the peaks evolve as well: at small $J/U$ they are almost perfectly asymmetric in frequency (leading to a vanishing integral, see Eq.~\eqref{eq:sum_rule_ReC}) while upon increasing $J/U$, when the system becomes more delocalized, this asymmetry disappears. Furthermore, also the strength of the peaks increases with $J/U$ (note the different scale in the panels) in a way that appears opposite to the peaks in the spectral function in the top panel. This is again consistent with the idea that upon entering in the delocalized regime the weight is transferred from the localized (on-site) modes to the delocalized (off-diagonal ones).

\begin{figure}[t]
    \centering
    \includegraphics[scale=0.45]{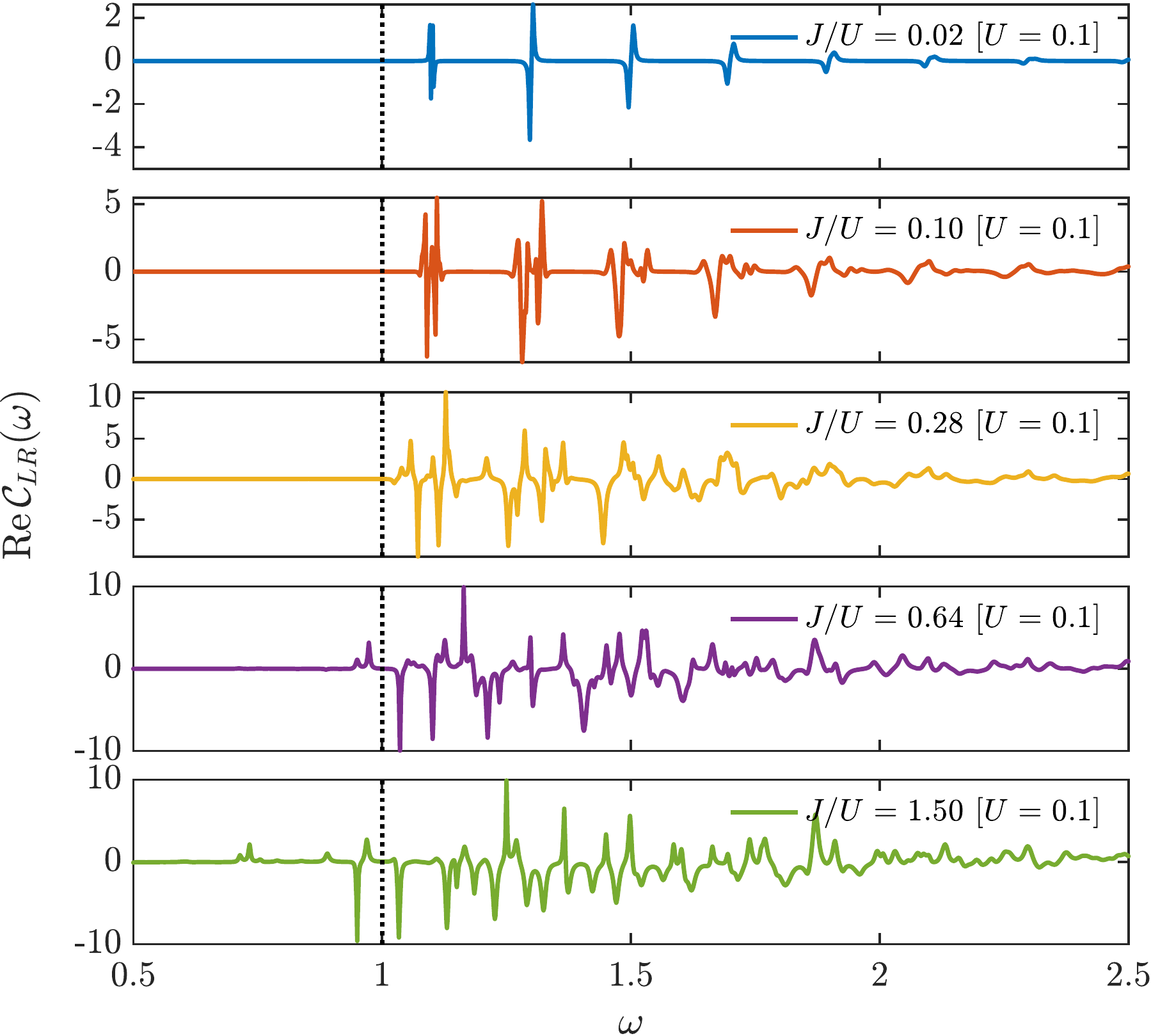}\\\medskip
    \includegraphics[scale=0.45]{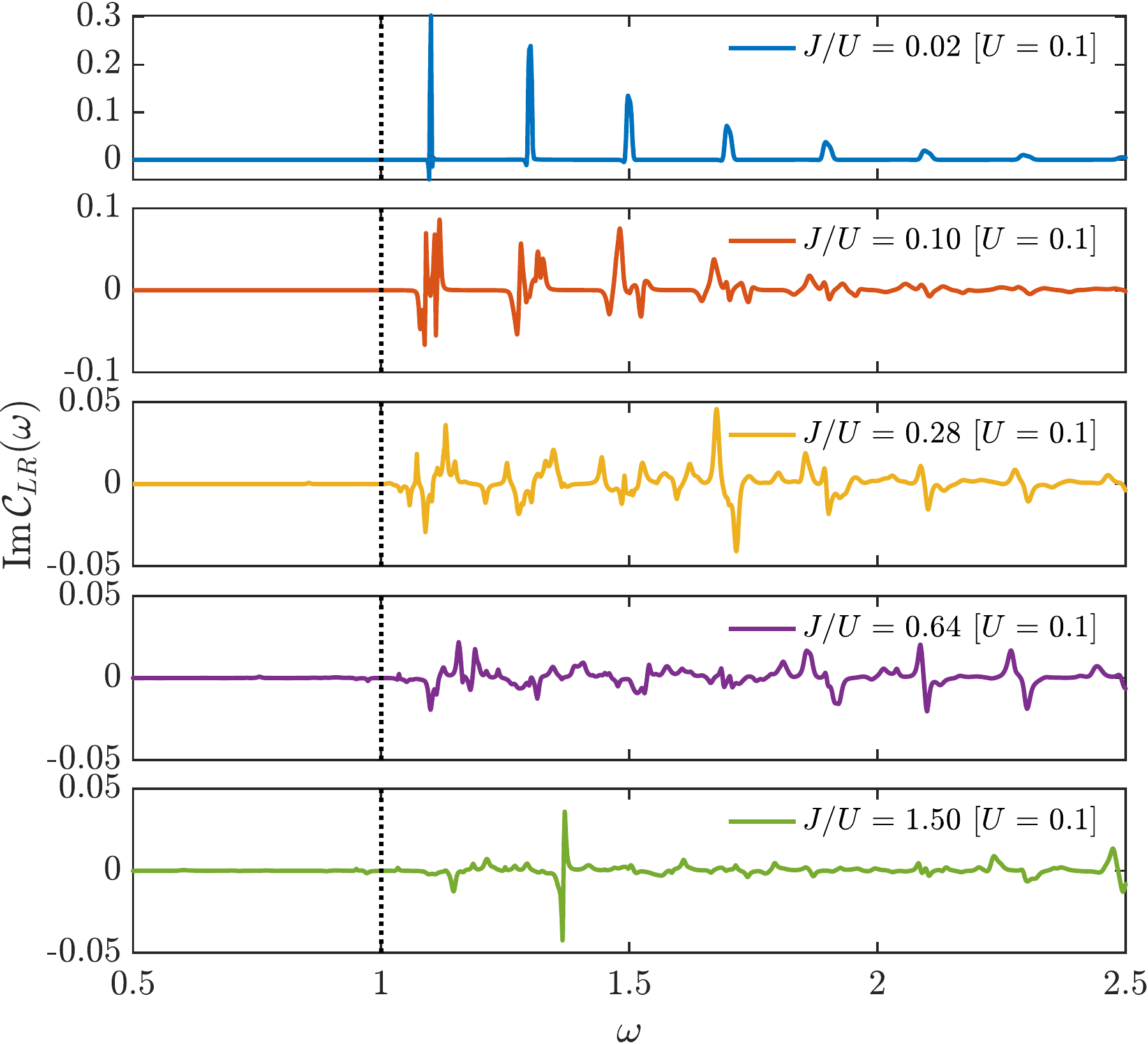}
    \caption{Cavity off-diagonal correlation function $C_{LR}(\omega)$ for different values of $J/U$ [$U=0.1$]. The cavities have a base frequency $\omega_0=1.0$ (vertical dotted line), while the effective losses are $\Gamma^{\mathrm{eff}}_L = 2.5 \times 10^{-4}$ and $\Gamma^{\mathrm{eff}}_R = 1 \times 10^{-4}$ and the pumping rates realizes uncoupled steady-state occupations equal to $\sim 3.3$ in the left cavity and $2$ in the right cavity.}
    \label{fig:multiGF_DP_Imb_continue}
\end{figure}

\subsection{Asymmetric Pump and Losses}
\label{sec:res_gf_asym}

We now move to discuss the case of asymmetric pump and losses, $\Delta P, \, \Delta\Gamma \neq 0$, resulting as we know in a non trivial stationary state density matrix (and finite imbalance, see Sec.~\ref{sec:res_quantum_ss}). A natural question is whether this different nonequilibrium protocol results in a qualitatively different behavior of the Green's functions. 

We start from the spectral functions, that we plot in Fig.~\ref{fig:multiGF_DP_Imb} for a fixed pump/loss asymmetry and different values of $J/U$. To highlight the comparison between the two cavities we plot the left and right spectral functions on a common frequency scale. While we see a similar structure of peaks evolving with $J/U$, as compared to the symmetric case of Fig.~\ref{fig:multiGF_Imb}, we also note an interesting dependence from the pump/loss asymmetry and the hopping. In particular, for small $J/U$ the right cavity spectral function (bottom panels) has slightly stronger peaks at low frequency than the left cavity one, reflecting the asymmetry in the pump/loss rates. As the hopping is increased and the excitations are delocalized in the dimer we see that this asymmetry in the left/right spectral functions decreases and for $J/U=1.50$ the two spectra are essentially the same and very close in shape to the symmetric one for the same value of $J/U$ (See Fig.~\ref{fig:multiGF_Imb}).

Then we consider the off-diagonal cavity correlation function, see Fig.~\ref{fig:multiGF_DP_Imb_continue}, that we study as a function of $J/U$. In the top panel we plot the real part, $\Real \mathcal{C}_{LR}(\omega)$, which shows a qualitative behavior very similar to the symmetric case shown in Fig.~\ref{fig:multiGF_Imb}, with anti-Lorentzian peaks which broaden and merge into a continuum at large $J/U$ indicating the increase in kinetic energy. On the other hand, an interesting difference appears in the imaginary part of the off-diagonal cavity correlation function, $\Imag \mathcal{C}_{LR}(\omega)$, which is now different from zero and shows a non-trivial dependence from $J/U$, with narrow peaks which broaden and merge into a continuum as $J/U$ is increased. 

\begin{figure}
    \centering
    \includegraphics[scale=0.49]{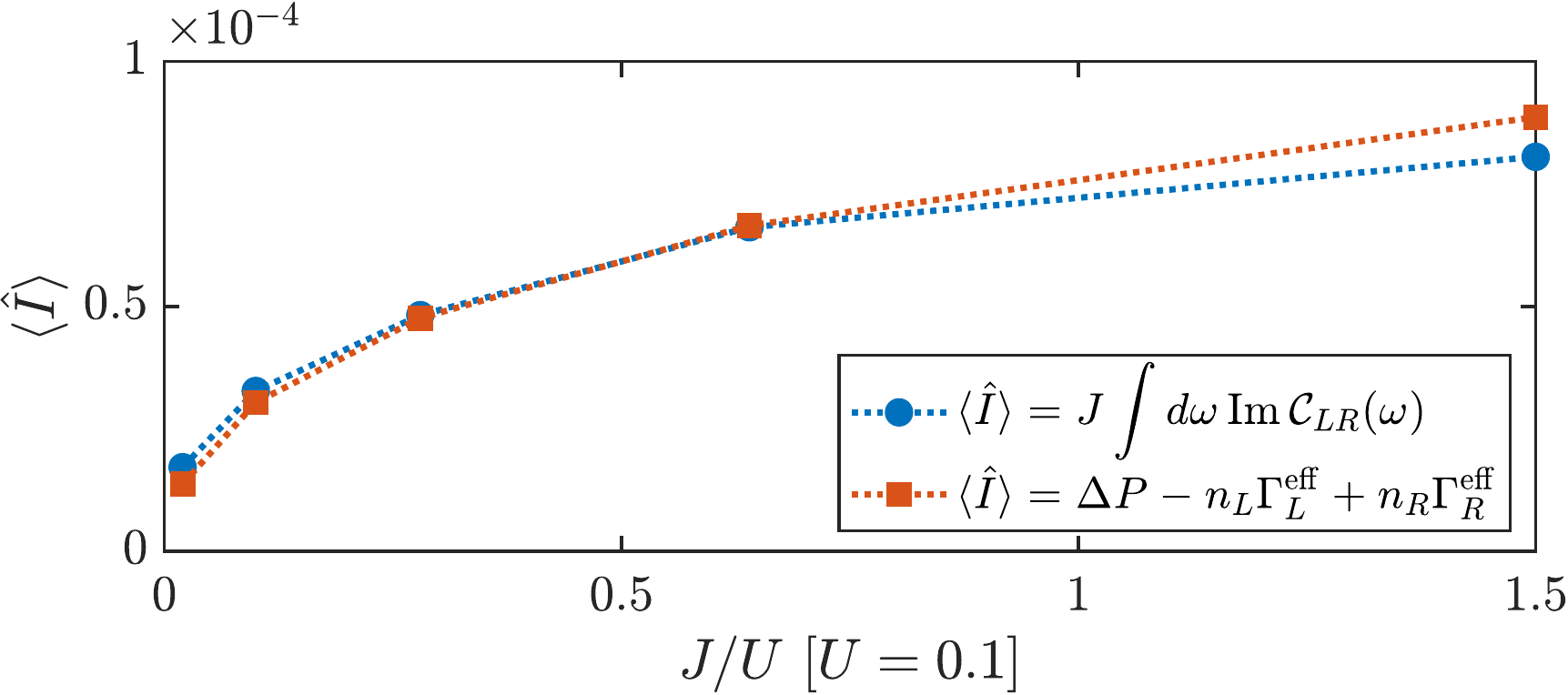}
    \caption{Current flowing in the dimer as obtained from \eqref{eq:sum_rule_ImC} and \eqref{eq:current_fom_DP}. The shown values of $J/U$ are the same ones used for the panels of Figs.~\ref{fig:multiGF_DP_Imb}--\ref{fig:multiGF_DP_Imb_continue}.}
    \label{fig:current_sum_rule}
\end{figure}

We can understand the origin of a finite imaginary part of the off-diagonal cavity correlation function by using the sum rule that relates the integral of $\Imag \mathcal{C}_{LR}(\omega)$ to the average current flowing from $L$ to $R$ (see Eq.~\eqref{eq:sum_rule_ImC} and Appendix~\ref{app:sumrules}). In the stationary state the average current is completely determined by the effective pump/loss rates $\Gamma^{\mathrm{eff}}_{L/R}=\Gamma_{L/R}-P_{L/R}$ and the stationary occupation $n_{L/R}$ through the relation
\begin{equation}
    \braket{\hat{I}}
    = \Delta P - n_L\Gamma^{\mathrm{eff}}_L + n_R\Gamma^{\mathrm{eff}}_R ,
    \label{eq:current_fom_DP}
\end{equation}
where $\Delta P$ is the pump asymmetry. We see that the right-hand side of this equation exactly vanishes in the symmetric case $\Delta P=0$, $\Gamma^{\mathrm{eff}}_L =\Gamma^{\mathrm{eff}}_R$ since as we know the occupations of the two cavities become equal ($n_L=n_R$). On the other hand for finite pump/loss asymmetry there is a finite current flowing from $L$ to $R$ and therefore an intra-dimer dissipation. This is interesting since the two cavities are only coupled by a coherent hopping coupling. As a result of this finite current and dissipation the imaginary part of the off-diagonal cavity correlation function has to be different from zero, both based on the sum-rule in Eq.~\eqref{eq:sum_rule_ImC} and on physical intuition. In Fig.~\ref{fig:current_sum_rule} we plot the average current versus $J/U$ and compare it with the integral over $\Imag \mathcal{C}_{LR}(\omega)$ to confirm the quantitative agreement. We also see that the overall current, although very small, increases with $J/U$, an effect which does not appear clearly from the shape of $\Imag \mathcal{C}_{LR}(\omega)$ in Fig.~\ref{fig:multiGF_DP_Imb_continue} but that is consistent with the idea that delocalization leads to more coherent exchange of excitations between the two cavities and therefore an increased current.

Finally, we have also considered the case of extreme pump/loss asymmetry, corresponding to the situation in which one of the two cavities is non-dissipative, i.e.\@ $\Gamma^{\mathrm{eff}}_R = P_R = 0$. Quite interestingly we have found that also in this case, as for perfectly symmetric rates, the current and the dissipative part of the off-diagonal cavity correlation function $\Imag \mathcal{C}_{LR}(\omega)$ are both zero, for any value of $J/U$. We can understand this result from a simple physical picture: in absence of a Markovian environment coupled to the right cavity the current flowing from left to right cannot be dissipated and bounces back, resulting in a zero net current. This can be also understood more formally, by looking at Eq.~(\ref{eq:current_fom_DP}) and by noting that for $\Gamma^{\mathrm{eff}}_R = P_R = 0$ this reduces to $\braket{\hat{I}} = \Gamma^{\mathrm{eff}}_L\left(n_{0L}-n_L\right)$. As we discuss in Appendix~\ref{app:steady_state_U0} in the limit $\Gamma^{\mathrm{eff}}_R = P_R = 0$ the left cavity occupation reduces to the one of an isolated left site coupled to Markovian pump and losses, i.e.\@ $n_{L}=n_{0L}$ resulting therefore in a vanishing current.

\section{Discussion}
\label{sec:discussion}

In this section we discuss our results on the BHD in the broader context of driven-dissipative phase transitons and comment more in detail on the experimental realization of our setup and our findings.

As for their closed system counterparts, dissipative phase transitions emerge sharply in the limit of thermodynamically large systems~\cite{Minganti2018,Landa2020}. In the open-system context this has been shown to arise when taking the large volume limit at fixed finite-density or in the limit of large photon numbers, correspondingly to a well defined classical limit. From this point of view it is not surprising that for our BHD the localization-delocalization transition that exists at the semiclassical level turns in a crossover in presence of quantum fluctuations. These are in fact particularly strong in the present case where the system size is finite and therefore the Liouvillian gap is non-vanishing. This does not exclude of course the presence of sharp nonequilibrium phase transitions for arrays of driven-dissipative cavities with incoherent pumping, as it has been indeed recently discussed~\cite{Biella2017,Scarlatella2018}.

As we mentioned in the introduction, the driven-dissipative BHD has been realized experimentally in a variety of quantum light-matter platforms. In circuit QED this can be done by considering the large detuning limit of two coupled Jaynes-Cummings (JC) units, which can be realized by capacitively coupling two resonators, each containing a transmon qubit. In this context the focus has been mostly on the case of coherently driven cavities, or of purely dissipative (lossy) dynamics, however an incoherent pump can be also engineered by weakly coupling each site of the dimer to a transmission line or to an incoherent noise~\cite{Hoffman2011}. In an actual experimental setting, the case of perfectly symmetric dimer is obviously more difficult to achieve due to local imperfections which introduce small disorder in the system. This however has been shown to remain controllable, particularly for small lattices~\cite{Underwood2012,Fitzpatrick2017}. Our results for the dynamics of the imbalance or its dependence from external parameters, as well as the Green's functions, can be directly measured experimentally. The former has been done in the context of a JC dimer through homodyne detection~\cite{Raftery2014}. The latter can be naturally addressed in a transmission/reflection experiment. Finally, in other light-matter platforms, such as semiconductor microcavities and photonic crystals, incoherent pumping is even more natural to realize, especially for lasing applications~\cite{Hamel2015}. We also mention the BHD is relevant for ultracold atomic gases experiments with double-well systems, and in this context controlled dissipative (incoherent) processes of pump and losses can be engineered by coupling to other bands.

\section{Conclusions}
\label{sec:conclusions}

In this article we have studied an open Bose-Hubbard dimer and investigated the possible signatures of a dissipative localization-delocalization transition or crossover, where upon tuning the ratio of coherent hopping versus local interaction an initial population imbalance is either trapped in one of the two cavities (self-trapping) or equally distributed across the dimer.

In the semiclassical limit of many photons per site, that we reviewed for completeness in Sec.~\ref{sec:semiclassical_dynamics_recap}, this transition is known to occur sharply for a purely conservative (Hamiltonian) dynamics and to remain present in the form of a short-time dynamical transition in presence of pumps and losses, while turning into a smooth crossover at long times.

In the full quantum regime the situation is particularly interesting since it is known that in absence of any asymmetry in the system parameters the stationary state density matrix is independent of any Hamiltonian coupling and only set by the pump and loss coefficients. To address therefore possible signatures of a dissipative self-trapping crossover one is forced to go beyond simple steady-state observables or to explicitly break the symmetry between the two cavities. To this extent we have exactly solved the problem by numerical diagonalization of the Lindbladian superoperator and obtained the stationary state, the full dissipative quantum dynamics and properties of the excitations on top of the stationary state, as encoded in the single-particle Green's functions, see Sec.~\ref{sec:methods} .

In Sec.~\ref{sec:quantum_time_dynamics} we have shown that the short-time dissipative dynamics shows clear signatures of a crossover between a localized behavior with finite residual imbalance and coherent oscillations leading to a vanishing imbalance, which can be accessed by either changing the ratio $J/U$ or the initial condition. On the other hand the long-times dynamics is largely controlled by the dissipative rates. In Sec.~\ref{sec:res_quantum_ss} we have shown that by breaking the symmetry of pump-loss rates between the two cavities one can induce a non-trivial stationary state and a finite imbalance which shows a smooth delocalization crossover upon increasing $J/U$. 

Finally, in Sec.~\ref{sec:res_gf} we have presented our results for the single particle Green's functions, in particular the spectral function and the cavity correlation function describing spectrum and occupation of the bosonic modes. These turn out to be sensitive probes of the Hamiltonian dynamics even in the fully symmetric case, where the delocalization crossover is signaled by the splitting of the lowest energy single-photon peak into bonding and anti-bonding modes as $J/U$ is increased. In presence of a finite pump-loss asymmetry we have shown that a finite current flows between the left and right cavities and this has direct consequences in the emergence of a non-vanishing imaginary part of the off-diagonal cavity correlation function.

The methodology discussed in this work, based on the exact diagonalization of a few-sites Lindbladian and on the computation of Green's functions, can be applied to different problems. Within the BHD it would be interesting to study the role of two-particle losses recently discussed in the context of the quantum Zeno effect~\cite{Misra1977,Peres1980,Itano1990,Syassen2008,Rossini2020}. Another future direction is the development of an exact diagonalization Lindblad impurity solver for Dynamical Mean Field Theory~\cite{Georges1996,Aoki2014,Arrigoni2013,Scarlatella2020}; in this scheme the DMFT self-consistent bath is approximated with a limited number of effective sites. In this respect we note that a two-site model turns out to share many similarities~\cite{Capone2002} with a minimal, yet reasonably accurate, implementation of the DMFT using a single site in the bath~\cite{Potthoff2001}. The rationale is simply that, in the dimer, one of the two sites plays the role of the self-consistent bath for the other.

\acknowledgments

This work was partially supported by the ANR grant ``NonEQuMat'' (ANR-19-CE47-0001) (M. Schirò) and by Italian MIUR through the PRIN2017 project CEnTral (Protocol Number 20172H2SC4).

\section*{Data Availability}

The data that support the findings of this study are available upon reasonable request from the authors.

\appendix

\section{Semiclassical Dynamics}
\label{app:semiclassical_dynamics}

The driven-dissipative Bose-Hubbard dimer can be also analyzed at a semiclassical level, by writing the Heisenberg equations for the cavity fields in \eqref{eq:dimer_hamiltonian} with pumping and losses as non-Hermitian terms and then taking the expectation values:
\begin{align}
    \dot{a}_{L} 
    &= -i \Big[ (\omega_L-U) + 2Ua_L^{\dagger}a_L  \Big] a_L - iJa_R - \Gamma_{L}^{\mathrm{eff}}a_L \nonumber \\
    \dot{a}_{R} 
    &= -i \Big[ (\omega_R-U) + 2Ua_R^{\dagger}a_R  \Big] a_R - iJa_L - \Gamma_{R}^{\mathrm{eff}}a_R \nonumber
\end{align}
where $\Gamma_{L/R}^{\mathrm{eff}} = \Gamma_{L/R} - P_{L/R}$ are the \emph{effective} loss rates, which for single-particle losses must always be positive.

By applying the transformation
\begin{equation}
    a_{L/R} \doteqdot \alpha_{L/R}e^{i\vartheta_{L/R}},
    \qquad
    \alpha_i,\vartheta_i \in \mathbb{R}
\end{equation}
one can then reduce the equations for the two complex numbers above into the following three equations for the real quantities $N = n_L + n_R$, $Z = n_L - n_R$ and $\phi = \vartheta_L - \vartheta_R$:
\begin{equation}
    \begin{cases}
        \displaystyle \dot{N} = -\left(\Gamma_{L}^{\mathrm{eff}}+\Gamma_{R}^{\mathrm{eff}}\right)N - \left(\Gamma_{L}^{\mathrm{eff}} - \Gamma_{R}^{\mathrm{eff}}\right)Z \\
        \begin{aligned}[c]%
            \displaystyle \dot{Z} =
            &-\left(\Gamma_{L}^{\mathrm{eff}}+\Gamma_{R}^{\mathrm{eff}}\right)Z - \left(\Gamma_{L}^{\mathrm{eff}} - \Gamma_{R}^{\mathrm{eff}}\right)N \\
            &- 2J\sqrt{N^2-Z^2}\sin(\phi)
        \end{aligned} \\
        \displaystyle \dot{\phi} = -\Delta\omega - 2UZ + 2J\frac{Z}{\sqrt{N^2-Z^2}}\cos\phi
    \end{cases}
\label{eq:BH_dimer_semiclassical_equations}
\end{equation}
where $\Delta\omega = \omega_L-\omega_R$.

\subsection{Closed System}

In the Hamiltonian case, with $\Delta\omega=0$ for simplicity, the equations reduce to
\begin{equation}
    \begin{cases}
        \displaystyle \dot{N} = 0 \qquad\Longrightarrow\qquad N = N_0 = \text{const}. \\
        \displaystyle \dot{Z} = - 2J\sqrt{N_0^2-Z^2}\sin(\phi) \\
        \displaystyle \dot{\phi} = - 2UZ + 2J\frac{Z}{\sqrt{N_0^2-Z^2}}\cos\phi
    \end{cases}
    \label{eq:BH_dimer_semiclassical_equations_hamiltonian}
\end{equation}
By using the fact that in a closed system the energy is conserved, the two remaining equations can be further reduced to a single equation for the macroscopic occupation imbalance:
\begin{equation}
    \dot{Z} = -2\sqrt{p(Z)}
    \label{eq:imbalance_equation_semi}
\end{equation}
where $p(Z)$ is a polynomial that can be factorized as
\begin{equation}
    p(Z) 
    = -\frac{U^2}{4}\left(Z^2-Z_0^2\right)\left(Z^2-Z_1^2\right),
\end{equation}
with $Z_0$ the initial imbalance and $Z_1$ equal to
\begin{equation}
    Z_1 = \sqrt{Z_0^2 + 4\left(\frac{J}{U}\right)\sqrt{N_0^2-Z_0^2} - 4\left(\frac{J}{U}\right)^2}.
\end{equation}

Being under a square root, the sign of $p(Z)$ is the real discriminant on the evolution of $Z$. In turn, the sign of $p(Z)$ is completely determined by $Z_1$ being real or imaginary (since $Z_0$ is real). If $Z_1$ is real then the polynomial is positive only in the region between $Z_0$ and $Z_1$ and in the region between $-Z_0$ and $-Z_1$, no matter whether $Z_1$ is greater or less than $Z_0$. If instead $Z_1$ is imaginary then the polynomial is positive only in the region between $-Z_0$ and $Z_0$.

The nature of $Z_1$ is in turn determined by the sign of the polynomial
\begin{equation}
    \left(\frac{J}{U}\right)^2 -\sqrt{N_0^2-Z_0^2}\left(\frac{J}{U}\right) -\frac{Z_0^2}{4}.
\end{equation}
If we assume that $J/U$ is positive, then the polynomial above provides a critical $\frac{J}{U}$, given in the main text in Eq.~\eqref{eq:J_over_U_semi_critical} that we rewrite here for simplicity
\begin{equation}
    \left(\frac{J}{U}\right)_{\mathrm{c}} = N_0\left( \frac{\sqrt{1-(Z_0/N_0)^2} + 1}{2} \right).
\end{equation}
For $\frac{J}{U} < \left(\frac{J}{U}\right)_{\mathrm{c}}$ $Z_1$ is real and therefore $Z(t)$ oscillates between $Z_0$ and $Z_1$; for $\frac{J}{U} > \left(\frac{J}{U}\right)_{\mathrm{c}}$ $Z_1$ is imaginary and therefore $Z(t)$ oscillates between $-Z_0$ and $Z_0$. Then $\left(\frac{J}{U}\right)_{\mathrm{c}}$, in this sense, can be interpreted as a critical value for a transition from a localized regime (low $J$) to a de-localized regime (high $J$).

\begin{figure}
    \centering
    \includegraphics[scale=0.49]{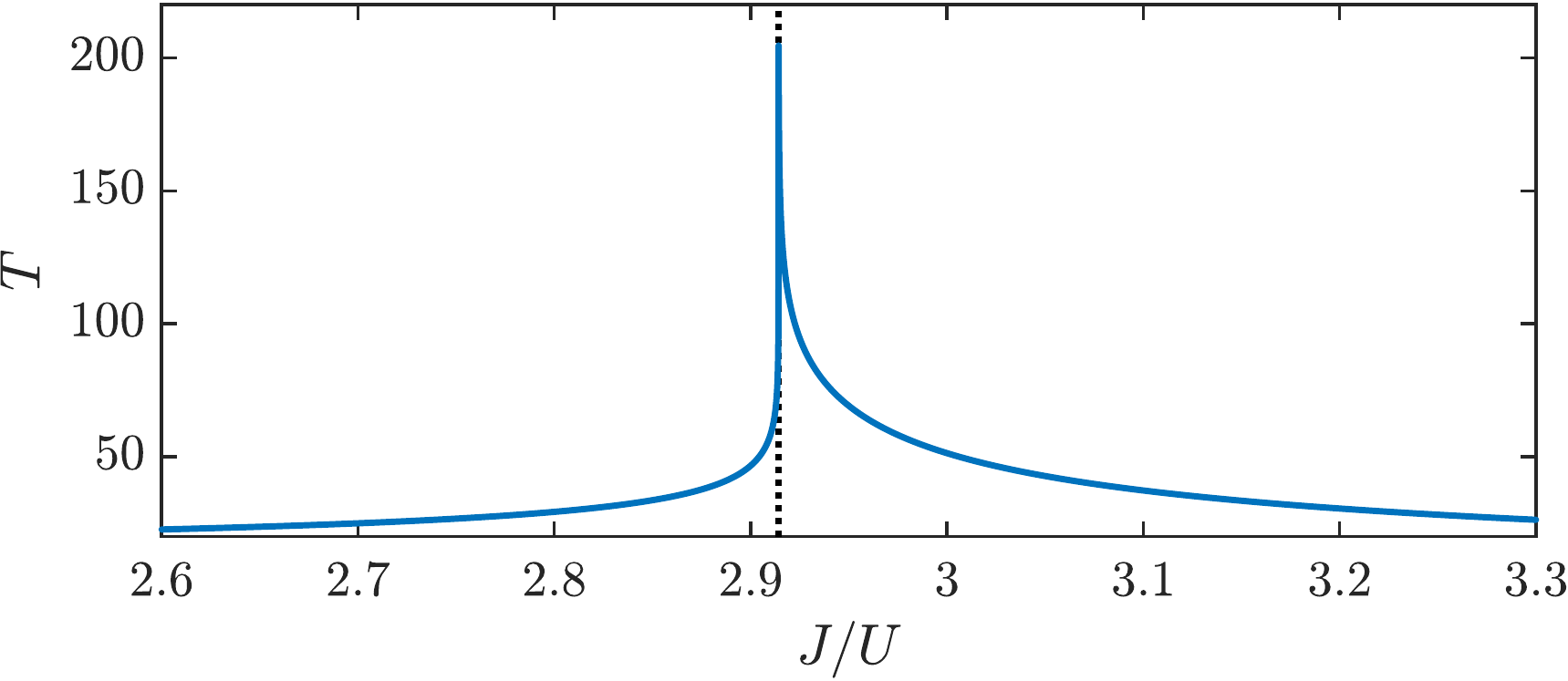}
    \caption{Oscillation period $T$ obtained from \eqref{eq:T_analytical_semi} as a function of $J/U$; the settings are the same as in Fig.~\ref{fig:Z_vs_time_semiclassical}.}
    \label{fig:semiclassical_T_divergence}
\end{figure}

This transition can also be seen through the divergence of the oscillation period at the critical point (Fig.~\ref{fig:semiclassical_T_divergence}), which can be analytically expressed as
\begin{equation}
    T = 
    \begin{cases}
    \frac{4K\left(\left(\frac{Z_0}{Z_1}\right)^2\right)}{U\sqrt{-Z_1^2}}  & \left(\frac{J}{U}\right) > \left(\frac{J}{U}\right)_c \\
    \left\lvert \frac{2\left[ K\left(\left(\frac{Z_0}{Z_1}\right)^2\right) - F\left(\sin^{-1}\left(\frac{Z_1}{Z_0}\right),\left(\frac{Z_0}{Z_1}\right)^2\right) \right]}{U\sqrt{-Z_1^2}} \right\rvert & \left(\frac{J}{U}\right) < \left(\frac{J}{U}\right)_c
    \end{cases}
    \label{eq:T_analytical_semi}
\end{equation}
where $F(\varphi,m) = \int_0^\varphi du \frac{1}{\sqrt{1-m^2\sin^2u}}$ and $K(m) = F(\frac{\pi}{2},m)$ are respectively the incomplete and the complete elliptic integral of the first kind.

The divergence is logarithmic, as one can infer by approximating the integral around the critical point ($Z_1 \to 0^+$). The fact that the period diverges, making the oscillations slower and slower, is a common signature of a phase transition and it's called \emph{critical slowing down}.

\subsection{Open System}

The question is now how much of the non-dissipative analysis done above survives in the presence of losses, at intermediate times. As we cannot go further with an analytical treatment, we have to go back to \eqref{eq:BH_dimer_semiclassical_equations} and solve the full system of equations.

Intuitively we expect to see a similar oscillatory behavior of $Z(t)$ in the dissipative case, though the mean value approaches zero at large enough times since, semiclassically, dissipative cavities decay to vacuum at the stationary state.

Indeed, you see that the presence of dissipation has the double effect of increasing the oscillation period and producing an overall decay of the occupation imbalance with time. But more interestingly, it stimulates a dynamical transition from the regime in which the imbalance oscillations are between $Z_0$ and $Z_1$ to a regime in which the imbalance oscillates around $0$.

\section{Analytical Quantum Results at \texorpdfstring{$U=0$}{U=0}}

\subsection{Green's Functions}
\label{app:greens_U0}

The Green's functions at $U=0$ can be obtained analytically via the Keldysh formalism. Here we start with the single-cavity Green's function and then extend to two coupled cavities.

\subsubsection{Single Cavity}
\label{app:exact_green_1C}

The retarded, advanced and Keldysh components of the Green's function are:
\begin{equation}
    G^{R/A}(\omega) = \frac{1}{\omega-\omega_0\pm i(\Gamma-P)},
\end{equation}
\begin{equation}
    G^{K}(\omega) = \frac{-2i(\Gamma+P)}{(\omega-\omega_0)^2 + (\Gamma-P)^2}.
    \label{eq:G_SC}
\end{equation}
The loss/pumping rates appear in couple as $\Gamma-P$, except for the Keldysh Green's function in which they also appear as $\Gamma+P$. This is a signature of the quantum nature of the system, encoded in the Keldysh Green's function, in the same way that it appears, for example, when adding quantum noise to a semiclassical treatment.

\subsubsection{Two coupled cavities}
\label{app:exact_green_2C}

In the case of two coupled cavities, we distinguish between left and right cavity with a subscript $L/R$. The uncoupled Green's functions, denoted with a subscript $0$, are the ones in \eqref{eq:G_SC} that we've found before for the single cavity, i.e.
\begin{align}
    G_{0i}^{R/A}(\omega) = \frac{1}{\Delta_i \pm i\Gamma_{-i}}, \qquad
    G_{0i}^{K}(\omega) = \frac{-2i\Gamma_{+i}}{\Delta_i^2 + \Gamma_{-i}^2}
    \label{eq:G0_2C}
\end{align}
where
\begin{align}
    \Delta_i &\doteqdot \omega - \omega_i, \qquad i = L,R \\
    \Gamma_{\pm i} &\doteqdot \Gamma_{i} \pm P_{i}, \qquad i = L,R
\end{align}

Then the Green's function components for the left cavity are
\begin{equation}
    G_{L}^R(\omega) = \frac{1}{\Delta_L+i\Gamma_{-L} - \frac{J^2}{\Delta_R+i\Gamma_{-R}}}
    \label{eq:G_R_L}
\end{equation}
\begin{equation}
    G_{L}^A(\omega) = (G_{L}^R(\omega))^*
    \label{eq:G_A_L}
\end{equation}
\begin{align}
    G_{L}^K(\omega)
    = -2i\Bigg[ \Gamma_{+L} + J^2\frac{\Gamma_{+R}}{\Gamma_{+R}^2 + \Gamma_{-R}^2} \Bigg] \left\lvert G_{L}^R(\omega) \right\rvert^2
    \label{eq:G_K_L}
\end{align}
and the corresponding Green's functions for the right cavity are obtained by simply replacing $L \to R$.

\subsection{Steady-State Properties}
\label{app:steady_state_U0}

\begin{figure}
    \centering
    \includegraphics[scale=0.49]{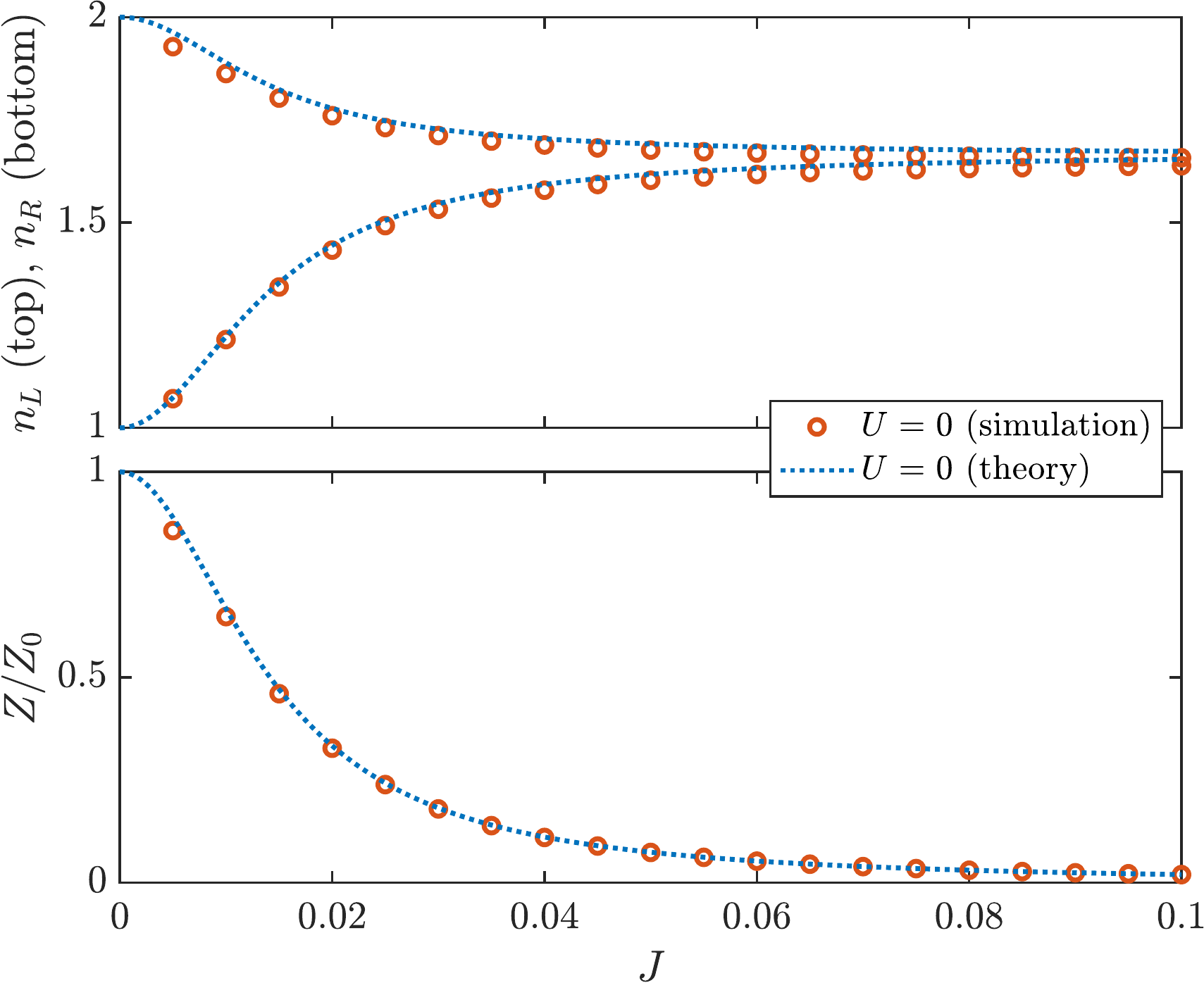}
    \caption{(Top) Steady-state cavity occupations as a function of $J$ for the dimer with loss coefficients $(\Gamma_L,\,\Gamma_R) = (6\times10^{-2},\,2\times10^{-2})$ and pump coefficients $(P_L,\,P_R) = (4\times10^{-2},\,1\times10^{-2})$ described in Sec.~\ref{sec:res_quantum_ss}, at different values of $U$. The dashed lines are the theoretical predictions for the $U=0$ case, calculated by combining the exact formula for the Keldysh Green's function \eqref{eq:G_K_L} and \eqref{eq:n_via_cavcorfun}. 
    (Bottom) Steady-state imbalance $Z = n_L - n_R$ corresponding to the occupations in the top panel.}
    \label{fig:2Cav_SS_0}
\end{figure}

The retarded Green's function of the left cavity can be also rewritten as
\begin{equation}
    G_{L}^R(\omega)
    = \frac{\Delta_{R}+i\Gamma_{-R}}{\Delta_{+}\Delta_{-} + i\left(\Delta_{L}\Gamma_{-R}+\Delta_{R}\Gamma_{-L}\right)}
\end{equation}
where $\Delta_{\pm} = \omega - \omega_{\pm}$ and
\begin{equation}
    \omega_{\pm} = \frac{\omega_L+\omega_R}{2} \pm \sqrt{\left(\frac{\omega_L-\omega_R}{2}\right)^2+J^2+\Gamma_{-L}\Gamma_{-R}}.
    \label{eq:omega_pm}
\end{equation}
Since the spectral function is proportional to the imaginary part of the retarded Green's function, this means that the frequency spectrum will be peaked around $\omega_+$ and $\omega_-$, and $J$ will just have the effect of increasing or decreasing the separation between these two peaks.

As for the occupations of the two cavities, they can be calculated via \eqref{eq:n_via_cavcorfun}. Analytical expressions can be easily obtained in some limiting cases. For example, if $\Gamma_{\pm R} = \Gamma_{\pm L}$, you obtain that $\int_{-\infty}^{+\infty} d\omega\,\mathcal{C}_L(\omega) = \Gamma_{+L} / \Gamma_{-L}$ and therefore
\begin{equation}
    n_L \equiv n_{0L} = \frac{P_{L}}{\Gamma_{L}-P_{L}}
    \label{eq:n_L_uncoupled}
\end{equation}
(and similarly for the right cavity), i.e.\@ the occupation of the cavities at the steady-state is equal to the occupation of the uncoupled cavities ($J=0$) and it's completely fixed by the pump/loss rates, no matter what the value of $J$ is. This is actually a special case of a result obtained in \cite{Lebreuilly2016}, showing that any number of cavities with the same incoherent pump/loss rates have a trivial steady state that does not depend on the details of their Hamiltonian, i.e.\@ in this case neither on $J$ nor on $U$. This means, in practice, that in order to have non-trivial physics at the steady state we must have, if not a loss imbalance between the two cavities, at least a \emph{pump imbalance}.

A more interesting case is the one at ``strong'' $J$, where ``strong'' means much bigger than at least all the loss coefficients. This time, we do not impose any prior condition on the pump/loss rates. If $\omega_L = \omega_R$ for simplicity, then the steady-state occupations become
\footnote{
    Note that the quantity $\Gamma_{-L/R}$ used in the quantum treatment has the same value of the semiclassical $\Gamma_{L/R}^{\mathrm{eff}}$.
    
    In addition, below the lasing threshold we can always parameterize $\Gamma_{L/R}$ and $P_{L/R}$ as 
    \begin{equation}
        \Gamma_{L/R} = \Gamma_{-L/R}\left( n_{0L/R} + 1 \right)
        \nonumber
    \end{equation}
    and
    \begin{equation}
        P_{L/R} = \Gamma_{-L/R}n_{0L/R}\,.
        \nonumber
    \end{equation}
}
\begin{equation}
    n_L \equiv n_R = \frac{\Gamma_{-L}n_{0L} + \Gamma_{-R}n_{0R}}{\Gamma_{-L} + \Gamma_{-R}},
    \label{eq:weighted_mean_occupations}
    \end{equation}
i.e., for strong enough coupling the occupation of the left and of the right cavities are equal and equal to a weighted average of their bare occupations. 

In particular, if the effective losses are equal ($\Gamma_{-L} = \Gamma_{-R}$), then
\begin{equation}
    n_L \equiv n_R = \frac{n_{0L} + n_{0R}}{2},
\end{equation}
i.e.\@ the steady-state occupation of the two cavities is exactly the mean between the bare occupations.

The $J=0$ and strong $J$ limits match our intuitive expectations, i.e.\@ that the occupations of the cavities, as a function of $J$, start from their uncoupled values and get closer and closer to each other as $J$ is increased, up to the point at which they match each other's value.

Another interesting limiting case is obtained if one of the cavities, say e.g.\@ the right one, has $\Gamma_{\pm R} = 0$. Then, for any $J$, we get
\begin{equation}
    n_L \equiv n_R \equiv n_{0L}.
\end{equation}
In this case the uncoupled occupation of the right cavity, $n_{0R}$, is formally ill-defined; however, it can be easily regularized by taking $P_R=0$ and $\Gamma_R = \varepsilon$, with $\varepsilon > 0$ arbitrarily small, for which $n_{0R} = 0$.

From a physical point of view, in this case the steady-state occupations in the system are fixed by the only available Markovian environments, i.e.\@ the ones attached to the left cavity, so the occupations become equal as soon as the two cavities are connected ($J>0$). For this reason, we expect this result to be valid at $U \neq 0$ as well.

\section{Sum-Rules and Particle Currents in the BHD}
\label{app:sumrules}

We start deriving the sum-rules for the off-diagonal correlation function defined in Eqs.~\eqref{eq:sum_rule_ReC}--\eqref{eq:sum_rule_ImC}. To this extent we note that, by our definitions in Eqs.~\eqref{eq:G_AB_RK_defs}--\eqref{eq:spectralA_L__cavcorfunC_L},
\begin{equation}
    \int_{-\infty}^{+\infty} d\omega\,e^{-i\omega t}\mathcal{C}_{LR}(\omega)
    = \braket{ \hat{a}_L(t)\hat{a}^{\dagger}_R+\hat{a}^{\dagger}_R\hat{a}_L(t) }
    \nonumber
\end{equation}
and that by taking the Hermitian conjugate we have
\begin{equation}
    \int_{-\infty}^{+\infty} d\omega\,e^{i\omega t}\mathcal{C}_{LR}^*(\omega)
    = \braket{ \hat{a}_R\hat{a}^{\dagger}_L(t)+\hat{a}^{\dagger}_L(t)\hat{a}_R }.
    \nonumber
\end{equation}
Taking the $t\rightarrow0^+$ limit and the sum/difference of the above two equations we obtain
\begin{equation}
    \int_{-\infty}^{+\infty} d\omega\,
    \left(\mathcal{C}_{LR}(\omega)+\mathcal{C}^*_{LR}(\omega)\right)
    = 2\braket{ \hat{a}^{\dagger}_L\hat{a}_R+\hat{a}^{\dagger}_R\hat{a}_L }
    \nonumber
\end{equation}
as well as
\begin{equation}
    \int_{-\infty}^{+\infty} d\omega\,
    \left(\mathcal{C}_{LR}(\omega)-\mathcal{C}^*_{LR}(\omega)\right)
    = 2\braket{ \hat{a}^{\dagger}_R\hat{a}_L-\hat{a}^{\dagger}_L\hat{a}_R },
    \nonumber
\end{equation}
from which the sum-rules quoted in the main text follow. 

We now relate the average stationary current across the dimer to the pump-loss asymmetry. To this extent we consider the BHD in Eq.~\eqref{eq:dimer_hamiltonian} and we start writing down the quantum equation of motion for the density of bosons in each site of the dimer, $n_{\alpha}(t)=\mathrm{Tr}\Big(\hat{\rho}(t) \hat{n}_{\alpha}\Big)$, with $\alpha=L/R$, which read
\begin{align}
    \frac{dn_L}{dt} 
    &= i\Braket{ \left[\hat{T},\hat{n}_L\right] }
    + 2\Big( P_L + n_L\left(P_L-\Gamma_L\right) \Big)\\
    \frac{dn_R}{dt} 
    &= i\Braket{ \left[\hat{T},\hat{n}_R\right] }
    + 2\Big( P_R + n_R\left(P_R-\Gamma_R\right) \Big)
\end{align}
where $\hat{T} = J\Big( \hat{a}_L^{\dagger}\hat{a}_R + \hat{a}_R^{\dagger}\hat{a}_L \Big)$ is the kinetic energy operator. The commutator gives
\begin{equation}
    \left[\hat{T},\hat{n}_L\right]
    = -\left[\hat{T},\hat{n}_R\right]
    = J\Big( \hat{a}_R^{\dagger}\hat{a}_L- \hat{a}_L^{\dagger}\hat{a}_R \Big)
    \equiv i\hat{I} \,.
\end{equation}
If we take the difference between the two equations we obtain for the dynamics of the imbalance $Z=n_L-n_R$ the result
\begin{equation}
    \frac{dZ}{dt}
    = -2\braket{\hat{I}}
    + 2\Big( \Delta P - n_L\Gamma^{\mathrm{eff}}_L + n_R\Gamma^{\mathrm{eff}}_R \Big)
\end{equation}
In the stationary state the right hand side goes to zero and we obtain
\begin{equation}
    \braket{\hat{I}}
    = \Delta P - n_L\Gamma^{\mathrm{eff}}_L + n_R\Gamma^{\mathrm{eff}}_R
\end{equation}
from which, using Eq.~\eqref{eq:n_L_uncoupled}, we immediately conclude that for symmetric pump and losses there is no average current between the two sites of the dimer and as a consequence, using Eq.~\eqref{eq:sum_rule_ImC}, the imaginary part of the off-diagonal cavity correlation function has vanishing integral.

\bibliographystyle{apsrev4-2}   
\typeout{}                      
\bibliography{references}

\end{document}